\documentclass[twocolumn,trackchanges]{aastex62}

\usepackage{amsmath}
\usepackage{longtable}

\newcommand{\angstrom}{\mbox{\AA}} 

\newcommand{\Msun}{\ensuremath{\unitstyle{M}_\odot}}
\newcommand{\Lsun}{\ensuremath{\unitstyle{L}_{\odot}}}
\newcommand{\Rsun}{\ensuremath{\unitstyle{R}_{\odot}}}

\newcommand{\code}[1]{\texttt{#1}}
\newcommand{\mesa}{\code{MESA}}
\newcommand{\MESA}{\mesa}

\newcommand{\stella}{\code{STELLA}}
\newcommand{\CMFGEN}{\code{CMFGEN}}
\newcommand{\STELLA}{\stella}

\newcommand{\Sedona}{\code{Sedona}}
\newcommand{\Kepler}{\code{Kepler}}

\RequirePackage{bm}

\newcommand{\dif}{\ensuremath{\mathrm{d}}}

\renewcommand{\Lsun}{L_\odot}
\renewcommand{\Msun}{M_\odot}
\renewcommand{\Rsun}{R_\odot}
\newcommand{\Mej}{M_{\rm ej}}
\newcommand{\alpham}{\alpha_{\rm MLT}}
\newcommand{\Menv}{M_{\rm env}}
\newcommand{\MZAMS}{M_{\rm ZAMS}}

\newcommand{\Mfb}{M_{\rm fallback}}
\newcommand{\Mf}{M_{\rm final}}
\newcommand{\Mten}{M_{10}}
\newcommand{\Mfifteen}{M_{15}}

\newcommand{\Rfh}{R_{500}}
\newcommand{\Eexp}{E_{\rm exp}}
\newcommand{\Edep}{E_{\rm dep}}
\newcommand{\Ebind}{E_{\rm bind}}
\newcommand{\Eint}{E}

\newcommand{\Etot}{E_{\rm tot}}
\newcommand{\Efoe}{E_{51}}
\newcommand{\Lfifty}{L_{50}}
\newcommand{\Lft}{L_{42}}

\newcommand{\Teff}{T_{\rm eff}}

\newcommand{\Lbol}{L_{\rm bol}}

\newcommand{\Lprog}{L_{\rm prog}}
\newcommand{\XHe}{X_{\rm He}}

\newcommand{\frad}{f_{\rm rad}}

\newcommand{\tni}{t_{\rm Ni}}
\newcommand{\tNi}{\tni}

\newcommand{\tpt}{t_{\rm p}}
\newcommand{\vFe}{v_{\rm Fe}}
\newcommand{\vPh}{v_{\rm Ph}}
\newcommand{\vSN}{v_{\rm SN}}

\newcommand{\vFef}{v_{\rm Fe,50}}
\newcommand{\vPhf}{v_{\rm Ph,50}}
\newcommand{\vPht}{v_{\rm Ph,15}}

\newcommand{\ooc}{\omega/\omega_{\rm crit}}

\newcommand{\tauIB}{\tau_{\rm IB}}

\newcommand{\tauS}{\tau_{\rm Sob}}
\newcommand{\Ni}{{^{56}{\rm Ni}}}
\newcommand{\Co}{{^{56}{\rm Co}}}
\newcommand{\Fe}{{^{56}{\rm Fe}}}
\newcommand{\NNi}{N_{\rm Ni}}
\newcommand{\MNi}{M_{\rm Ni}}
\newcommand{\foe}{10^{51}\,{\rm ergs}}
\newcommand{\Feline}{Fe~II~5169\angstrom~line}
\newcommand{\FeA}{Fe~II~5169\angstrom}
\newcommand{\tfit}{t_{\rm fit}}
\newcommand{\ttau}{t_{\tau=10}}

\newcommand{\ET}{ET}
\newcommand{\ETff}{\ET_{55}}
\newcommand{\etaNi}{\eta_{\rm Ni}}
\newcommand{\QNi}{Q_{\rm Ni}}
\newcommand{\tday}{t_{\rm d}}
\newcommand{\tptwo}{t_{\rm p, 2}}
\newcommand{\tnone}{t_{0}}
\newcommand{\te}{t_{\rm e}}
\newcommand{\tSB}{t_{\rm SB}}
\newcommand{\tsb}{\tSB}
\newcommand{\days}{{\rm d}}

\newcommand{\appropto}{\mathrel{\vcenter{
		\offinterlineskip\halign{\hfil$##$\cr
	\propto\cr\noalign{\kern2pt}\sim\cr\noalign{\kern-2pt}}}}}

\newcommand{\mesafour}{MESA~IV} 

\newlength{\apjcolwidth}
\newlength{\figwidth}
\newlength{\doublewide}
\begin{document}
\title{Inferring Explosion Properties from Type II-Plateau Supernova Light Curves}

\author[0000-0003-1012-3031]{Jared A. Goldberg}
\affiliation{Department of Physics, University of California, Santa Barbara, CA 93106, USA}

\author{Lars Bildsten}
\affiliation{Department of Physics, University of California, Santa Barbara, CA 93106, USA}
\affiliation{Kavli Institute for Theoretical Physics, University of California, Santa Barbara, CA 93106, USA}

\author{Bill Paxton}
\affiliation{Kavli Institute for Theoretical Physics, University of California, Santa Barbara, CA 93106, USA}

\correspondingauthor{J. A. Goldberg}
\email{goldberg@physics.ucsb.edu}

\begin{abstract}
We present advances in modeling Type IIP supernovae using \texttt{MESA} for evolution to shock 
breakout coupled with \texttt{STELLA} for generating light and radial velocity curves.
Explosion models and synthetic light curves can be used to translate observable 
properties of supernovae (such as the luminosity at day 50 and the duration of the plateau, 
as well as the observable quantity $ET$, defined as the time-weighted integrated 
luminosity that would have been generated if there was no ${\rm ^{56}Ni}$ in the ejecta) 
into families of explosions which produce the same light curve and velocities 
on the plateau. These predicted families of explosions provide a useful guide 
towards modeling observed SNe, and can constrain explosion properties when coupled 
with other observational or theoretical constraints.
For an observed supernova with a measured ${\rm ^{56}Ni}$ mass, breaking the degeneracies 
within these families of explosions (ejecta mass, explosion energy, and progenitor radius) 
requires independent knowledge of one parameter. We expect the most common case 
to be a progenitor radius measurement for a nearby supernova. We show that ejecta 
velocities inferred from the Fe II 5169 \AA\ line measured during the majority of 
the plateau phase provide little additional information about explosion 
characteristics. Only during the initial shock cooling phase can photospheric 
velocity measurements potentially aid in unraveling light curve degeneracies. 
\end{abstract}


\section{Introduction\label{sec:INTRODUCTION}}

Through an expanding network of ground- and space-based telescopes, 
the astrophysical community has an unprecedented ability to probe transient events. 
Along with a host of facilities, 
such as the All Sky Automated Survey for Supernovae (ASAS-SN; \citealt{ASSASN2017}), 
the Las Cumbres Observatory \citep{LCO2013} is building the largest 
set of data ever collected on all nearby supernova (SN) events. Some SNe discovered have 
known progenitors in distant galaxies \citep{Smartt2009}. 
And the data are improving --- The Zwicky Transient Facility (ZTF; \citealt{ZTF2019})
has begun discovering multiple SNe on a nightly basis, and the Large Synoptic Survey Telescope (LSST; \citealt{LSST2009}) will 
revolutionize time-domain astronomy with repeated nightly imaging of the entire sky with outstanding spatial 
resolution. 

In this paper we focus on Type IIP SNe, core-collapse events of dying massive stars ($M>10\Msun$) 
which yield distinctive light curves that plateau over a period of $\approx$100 days.
The duration and brightness of these light curves reflect the progenitor's 
radius ($R$), ejected mass ($\Mej$), energy of the explosion ($\Eexp$), and $\Ni$ mass ($\MNi$).
Inferring these properties from the observations has broad applications. 
Extracting progenitor information from SN observations
could lend insight into which stars explode as SNe and which collapse directly into black holes. 
It would also have implications for the missing red supergiant (RSG) problem identified by 
\citet{Smartt2009} and updated by \citet{Smartt2015}, whereby Type II SNe with known progenitors
seem to come from explosions of RSGs with initial masses of $\MZAMS < 17\Msun$, whereas evolutionary
models have a cutoff mass of around $30\Msun$.

Our understanding has benefitted from 3-dimensional modeling of light curves and spectroscopic data for 
specific Type IIP events, such as the work of \citet{Wongwathanarat2015} and \citet{Utrobin2017}, as well as 3D 
simulations which probe specific regions of parameter space of these SNe (e.g. \citealt{Burrows2019}). 
Although 3D models are incredibly useful for describing specific systems and probing 
specific regions of the possible parameter space of progenitors and their explosions, 
substantial effort is required to estimate the parameters of a single observed explosion. 
The computational demand for individual 3D calculations presents a challenge
for probing the parameter space of possible progenitor models for a large ensemble of explosions. 

Here, we utilize the open-source 1-dimensional stellar evolution software instrument, 
Modules for Experiments in Stellar Astrophysics 
(\MESA; \citealt{Paxton2011,Paxton2013,Paxton2015,Paxton2018,Paxton2019}), 
to model an ensemble of Type IIP SN progenitors, interfacing with the radiative transfer code \STELLA\ 
\citep{Blinnikov1998,Blinnikov2004,Baklanov2005,Blinnikov2006} to simulate their light curves and photospheric 
evolution. We include the effects of the \citet{Duffell2016} prescription for mixing via the Rayleigh-Taylor 
Instability, which allows for significant mixing of important chemical species such as $\Ni$, and yields a
more realistic density and temperature profile in the ejecta at shock breakout 
(\citealt{Paxton2018}, MESA IV).

The increasing abundance of data has led to a new approach to understanding Type IIP progenitors and 
explosions in an ensemble fashion. \citet{Pejcha2015b,Pejcha2015a}, and \citet{Muller2017} 
took such an approach, characterizing a total of 38 Type IIP SNe by their luminosity
and duration of the plateau, as well as the velocity at day 50 
as inferred via the Fe II 5169 \angstrom\ line. By fitting these three measurements 
to the analytics of \citet{Popov1993}\footnote{See also \citealt{Sukhbold2016}'s 
update to the \Kepler\ results of \citealt{Kasen2009}, which find similar scalings.} 
and early numerics of \citet{Litvinova1983},
these authors inferred $\Mej$, $\Eexp$ and $R$ from these observables.

To this end, we show that \MESA+\STELLA\ reproduces a scaling for plateau luminosity at day 50, $\Lfifty$, 
similar to that of \citet{Popov1993}, and we introduce new best-fit scaling laws for 
$\Lfifty$ and for the duration of the plateau $\tpt$ in the limit of $\Ni$-rich events. 
We also discuss the relationship between our model properties and the observable $\ET$, 
the time-weighted integrated luminosity that would have 
been generated if there was no ${\rm ^{56}Ni}$ in the ejecta
\citep{Shussman2016, Nakar2016}, and show how $\ET$ can also be used to provide 
similar constraints on explosion properties. 
As an observable, $\ET$ is defined by Equations \eqref{eq:ETeq} and \eqref{eq:QNi}. 
Additionally, we show 
that the measured velocity at day 50 from the \Feline\ does not scale with ejecta mass and explosion energy 
in the way assumed by \citet{Popov1993}. Rather, as found observationally by \citet{Hamuy2003} and explained 
by \citet{Kasen2009}, agreement in $\Lfifty$ entails agreement in velocities measured near the photosphere at day 50 
(as we show in Figures \ref{fig:v50vsL50} and \ref{fig:99emMatching}). 

As our work was being completed, \citet{Dessart2019} submitted a paper that also highlights 
the non-uniqueness of light curve modeling for varied progenitor masses due to core size and mass loss 
due to winds. Here we additionally highlight the non-uniqueness of light curve modeling even for varied 
ejecta mass. As such, our calculated scaling relationships yield families of explosions with varied $R$, $\Mej$, 
and $\Eexp$ which could produce comparable light curves and similar observed \Feline\ velocities 
(e.g. see Figures \ref{fig:WedgeDiagram1} and \ref{fig:DegenerateLCs}).
Given an independent measurement of the progenitor $R$, along with a bolometric 
light curve and an observed nickel mass ($\MNi$) extracted from the tail, one can directly constrain 
$\Mej$ and $\Eexp$. Otherwise, these families of explosions can be used as a starting point to 
guide further detailed, possibly 3D, modeling for observed events. 

\section{Our Models\label{sec:MODELS}}

Our modeling takes place in three steps. First, we construct a suite of core-collapse supernova progenitor
models through the Si burning phase using \MESA\ 
following the \verb|example_make_pre_ccsn| test case, described in detail in \citealt{Paxton2018} (MESA IV). 
Second, we load a given progenitor model at core infall, excise the core (as described in section 6.1 of MESA IV), 
inject energy and Ni, and evolve the model until it approaches shock breakout. This closely follows the 
\verb|example_ccsn_IIp| test case. Third, to calculate photospheric evolution and light curves after shock breakout, 
we use the shock breakout profile produced in the second step as input into the public distribution of \STELLA\ included 
within \MESA, and run until day 175. At the end of the \STELLA\ run, a post-processing script produces data for comparison to 
observational results (specifically bolometric light curves and \Feline\ velocities as described in \mesafour).

In order to create a diversity of progenitor characteristics, we chose models with variations
in initial mass $\MZAMS$, core overshooting $f_{\rm ov}$ and $f_{\rm 0, ov}$, 
convective efficiency $\alpham$ in the hydrogen envelope, wind efficiency $\eta_{\rm wind}$, 
modest surface rotation $\omega/\omega_{\rm crit}$, and initial metallicity $Z$. 
This study concerns itself especially with achieving diversity in the ejecta mass $\Mej$ 
by means of the final mass at the time of explosion $\Mf$, and the radius $R$ at the time of the explosion. 
Table \ref{tab:progenitors} lists physical characteristics of all progenitor models utilized in this paper 
with $\Lprog=$ the stellar luminosity just prior to explosion. 
Our naming convention is determined by the ejecta mass and radius at shock 
breakout, M$<$$\Mej$$>$\_R$<$$R$$>$. For our sample of Type IIP SNe models, we use three progenitor models 
from \mesafour, the 99em\_19, 99em\_16, and 05cs models, renamed M16.3\_R608, 
M12.9\_R766, and M11.3\_R541, respectively. Additionally, we create three new models 
using \MESA\ revision 10398 to capture different regions of 
parameter space. We created a model with the same input parameters as 99em\_19,
here named M15.7\_R800. In order to explore a diversity of radii for similar
parameters, we also created M15.0\_R1140, a model with nearly identical input 
to M15.7\_R800, except for reduced efficiency of convective mixing $\alpham=2.0$ 
(the default value is $\alpham = 3.0$) to create a more radially extended star with 
otherwise similar properties. Finally, in order to include smaller progenitor radii 
and mass in our suite, we created M9.3\_R433, which has the same progenitor parameters 
as the 12A-like progenitor model from MESA IV, except greater overshooting 
$f_{\rm ov} = 0.01$. These ``standard suite" models are denoted by a *
in Table \ref{tab:progenitors}. All models are solar metallicity, except the 05cs-like progenitor 
from MESA IV, M11.3\_R541, which has metallicity $Z=0.006$.

Beyond this standard suite, we construct M20.8\_R969, a 
$\MZAMS=25\Msun$ non-rotating model with no overshooting and
wind efficiency $\eta_{\rm wind}=0.4$, which has a very tightly bound core and leads to significant 
fallback at energies $\Eexp<2\times\foe$ (see also Appendix \ref{sec:FALLBACK}). 
Additionally, to highlight the families of explosions which produce comparable light curves 
(see Section \ref{sec:PREDICTIONS}), we construct three progenitor models which, when exploded with the 
proper explosion energy, all produce light curves similar to that of our M12.9\_R766 model exploded with $0.6\times\foe$ 
and $\MNi=0.045\Msun$. M9.8\_R909 was $\MZAMS=$13.7$\Msun$ with a final mass of 11.4$\Msun$, created with 
overshooting $f_{\rm ov} = 0.016$, $f_{\rm 0,ov}=0.006$, initial rotation $(v/v_c)_{\rm ZAMS}=0.2$, 
wind efficiency $\eta_{\rm wind}=1.0$,  and $\alpham=2.0$. 
M10.2\_R848 was $\MZAMS=13.5\Msun$ with a final mass of 12.0$\Msun$, which was created with 
overshooting $f_{\rm ov} = 0.01$, $f_{\rm 0,ov}=0.004$,
initial rotation $(v/v_c)_{\rm ZAMS}=0.2$, wind efficiency $\eta_{\rm wind}=0.8$,  and $\alpham=2.0$. 
M17.8\_R587 was $\MZAMS=$20.0$\Msun$ with a final mass of 19.41$\Msun$, which was created with 
no overshooting, no rotation,  wind efficiency $\eta_{\rm wind}=0.2$, and $\alpham=3.5$.

\begin{table*}
\caption{Properties of our RSG progenitor models at the time of the explosion. Models indicated by a * are part of 
our standard suite, upon which the majority of our analysis is based. $M_{\rm{c}, i}$ is the mass of the excised core, 
and $M_{\rm He\ core}$ is the He core mass in the pre-explosion model. Names are determined by the ejecta 
mass and radius of each progenitor model at shock breakout, 
M$<$$\Mej$$>$\_R$<$$R$$>$.}
\centering
\singlespace
\begin{tabular}{cccccccccccc}
\hline\hline
model          & $M_{\rm ZAMS}$ & $\Mf$     & $M_{\rm{c}, i}$  & $M_{\rm He\ core}$ & $\Mej$    & $\Menv$  &  $\ooc$  & $|\Ebind|$        & $\Teff$  & log($\Lprog/\Lsun$) & $R$       \\
               & [$\Msun$]      & [$\Msun$] & [$\Msun$]        & [$\Msun$]          & [$\Msun$] & [$\Msun$]&          & [$10^{50}$ erg]  & [K]      &                     & [$\Rsun$] \\\hline
M9.3\_R433*    & 11.8           & 10.71     & 1.44             & 3.58               &  9.28     & 7.13     &  0.2     & 1.39              & 4370     & 4.79                & 433  \\
M11.3\_R541*   & 13.0           & 12.86     & 1.57             & 4.22               & 11.29     & 8.65     &  0.0     & 2.65              & 4280     & 4.95                & 541  \\
M12.9\_R766*   & 16.0           & 14.46     & 1.58             & 5.44               & 12.88     & 9.02     &  0.2     & 4.45              & 3960     & 5.11                & 766  \\
M16.3\_R608*   & 19.0           & 17.79     & 1.51             & 5.72               & 16.29     & 12.07    &  0.2     & 1.29              & 4490     & 5.13                & 608  \\
M15.7\_R800*   & 19.0           & 17.33     & 1.66             & 6.83               & 15.67     & 10.50    &  0.2     & 2.39              & 4040     & 5.18                & 800  \\
M15.0\_R1140*  & 19.0           & 16.77     & 1.78             & 7.55               & 14.99     & 9.22     &  0.2     & 3.31              & 3660     & 5.32                & 1140 \\ \hline
M20.8\_R969    & 25.0           & 22.28     & 1.77             & 8.85               & 20.76     & 13.43    &  0.0     & 8.45              & 4870     & 5.68                & 969  \\
M9.8\_R909     & 13.7           & 11.36     & 1.60             & 7.75               & 9.8       & 3.61     &  0.2     & 1.81              & 2380     & 4.99                & 909  \\
M10.2\_R848    & 13.5           & 11.99     & 1.77             & 4.24               & 10.22     & 7.75     &  0.2     & 1.92              & 3510     & 5.13                & 848  \\
M17.8\_R587    & 20.0           & 19.41     & 1.62             & 7.23               & 17.79     & 12.18    &  0.0     & 2.41              & 5480     & 5.44                & 587  \\
\hline   
\label{tab:progenitors} 
\end{tabular} 
\end{table*}

During the explosion phase, which we carry out using \MESA\ revision 10925 to include an updated treatment of fallback 
(see Appendix \ref{sec:FALLBACK}), 
we vary the total energy of the stellar model at the time of 
explosion ($\Etot$) from $2\times10^{50}$~ergs to $2\times\foe$, 
with 0.2, 0.3, 0.4, 0.5, 0.6, 0.7, 0.8, 1.0, 1.2, 1.4, 1.6, and 2.0 $\times\foe$. 
These models are significantly impacted by the \citet{Duffell2016} prescription for mixing via the Rayleigh-Taylor instability, 
which smooths out the density profile and leads to the mixing of H deep into the interior of the ejecta 
and $\Ni$ out towards the outer ejecta (see MESA IV). 
We use the RTI coefficient $D_\mathcal{R}=3.0$. For a further exploration of the impact of changing the 
strength of RTI-driven mixing on ejecta and light curve evolution, see the work of P. Duffell et al. (2019, in preparation).

At the handoff between \MESA\ and \STELLA, we initialize \STELLA\ with 400 zones and 40 frequency bins, and an error tolerance 
0.001 for the Gear-Brayton method \citep{Gear1971, Brayton1972}, which leads to converged models. We also 
rescale the abundance profile of $\Ni$ and $\Co$ to match a specified total Nickel mass $\MNi$. This resets the Nickel 
decay clock to the time of shock breakout. 
We consider $\Ni$ masses of $\MNi/\Msun=$0.0, 0.015, 0.03, 0.045, 0.06, and 0.075; the impact of $\Ni$ in our models 
is discussed in detail in Section \ref{sec:DURATIONS}. 
As most of the mixing is accounted for by Duffell RTI, we only employ modest boxcar smoothing of abundance profiles 
at handoff as recommended in MESA IV, using 3 boxcar passes with a width of 0.8 $\Msun$. 
Additionally, as described in \citealt{Paxton2019} and Appendix \ref{sec:FALLBACK} here,
we use a minimum innermost velocity cut of material moving slower than 500~km~s$^{-1}$ to prevent numerical artifacts 
in \STELLA\ caused during interactions between reverse shocks and slow-moving material near \STELLA's inner boundary.
This study concerns itself with intrinsic properties of the SNe and their progenitors, determined primarily by 
quantities on the plateau, and therefore we do not include circumstellar material (CSM) in \STELLA.

\subsection{Estimating Fallback}
\begin{figure}
\centering
\includegraphics[width=\columnwidth]{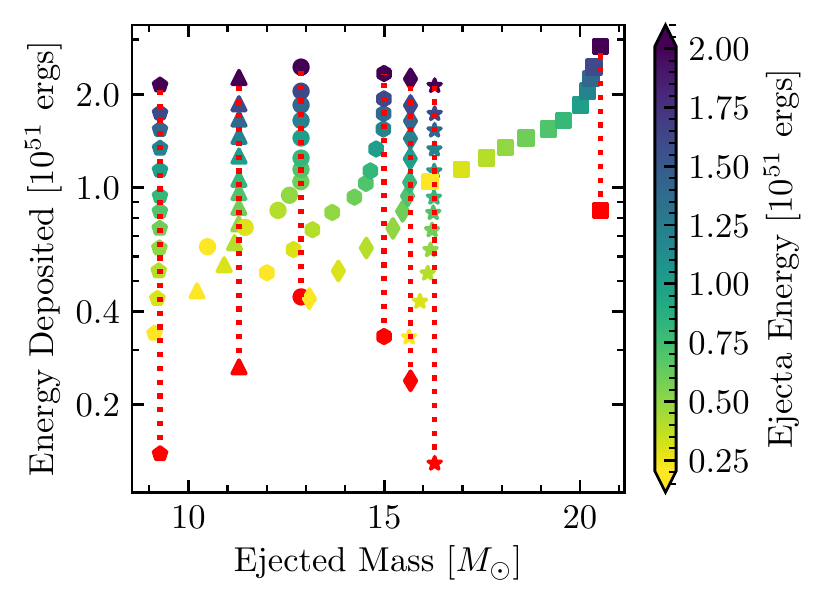} 
\caption{Energy deposited in the explosion versus ejected mass for our standard suite, as well as for the M20.8\_R969 model,
all exploded at 12 different explosion energies. Marker shapes correspond to different 
initial progenitor models as follows --- pentagon: M9.3\_ R433; triangle: M11.3\_R541; 
circle: M12.9\_R766; hexagon: M15.0\_R1140; diamond: M15.7\_R800; star: M16.3\_R608; 
square: M20.8\_R969. Explosions 
which would have $\Etot=0$ (corresponding to $\Edep=|\Ebind|$) are shown as red points whose x-coordinate 
is determined by the same model assuming no fallback. The red dotted line serves 
as a visual guide, indicating explosions with no fallback for each progenitor model. 
Color corresponds to the total energy of the ejecta just after the explosion $\Etot$.}
\label{fig:NiceFallback}
\end{figure}

Even when the total energy of a stellar model is greater than zero (i.e. the star is unbound),
it is possible for some of the mass which does not collapse into the 
initial remnant object to become bound and fall back onto the central object, which we define as $\Mfb$. This typically
occurs as a result of inward-propagating shock waves generated at compositional boundaries within the ejecta.  
The relationship between progenitor binding energy, explosion energy, and fallback can be seen in Figure \ref{fig:NiceFallback}, 
which shows the final mass of our models versus the total energy deposited $\Edep$, which is equal to the total energy
of the model after the explosion $\Etot$ plus the magnitude of the total energy of the bound progenitor model 
at the time of explosion $\Ebind$. Fallback is particularly common in explosions where the explosion energy 
is not significantly larger than the binding energy of the model at the time of explosion. In general, 
more tightly bound models require larger total final energies to unbind the entirety of the potential ejecta. 

The proper treatment of fallback in 1D simulations remains an open question because of complexities 
such as the interaction between accretion-powered luminosity and the inner boundary of the explosion models.  
In \MESA, the current implementation of fallback is effective as a computationally robust approximation that 
allows experimentation, but it should not be viewed as an accurate model of the physical processes at work.   
Consequently we restrict our study to models with little fallback material: $\Mfb<0.4\Msun$. 
The models which survive this cut are shown in figure \ref{fig:ParameterSpace}.  For a full description of 
our treatment of fallback, see Appendix~\ref{sec:FALLBACK}. 

\begin{figure}
\centering
\includegraphics{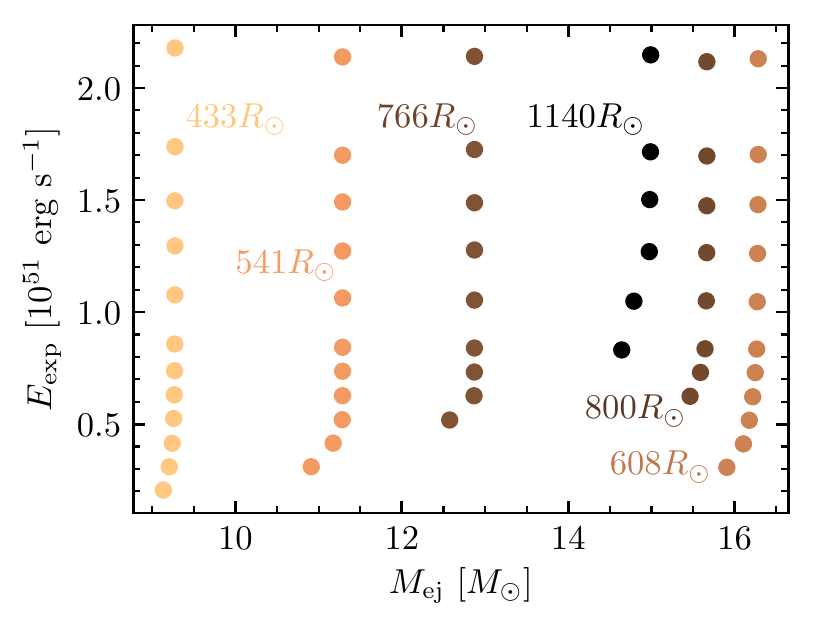} 
\caption{Ejecta masses and explosion energies as determined by the kinetic energy
at day 50 in \STELLA\ considered as a part of our standard suite, 
with fallback masses $\Mfb<0.4\Msun$. Lighter colors correspond to 
smaller progenitor radii, which are labeled.}
\label{fig:ParameterSpace}
\end{figure}

\section{Analytic Expectations\label{sec:PEDAGOGY}}

The luminosity of a Type IIP SN is, approximately, powered by shock cooling due to expansion out to around 20 days 
(the ``shock cooling phase"), then Hydrogen recombination until around 100 days (the ``plateau phase"), 
and the radioactive decay chain of $\Ni\rightarrow\Co\rightarrow\Fe$ beyond that (the ``Nickel tail"). 

The expansion time of the SN ejecta is expressed as $\te \approx R/\vSN$, 
where $R$ is the radius of the star at the time of the explosion, and the velocity $\vSN$ is defined 
by the mass of the ejecta $\Mej$ and kinetic energy of the ejecta at infinity $\Eexp = \Mej\,\vSN^2/2$.\footnote{During 
the the homologous phase, the kinetic energy of the ejecta is approximately equivalent to the total energy of 
the explosion, since radiation accounts only for a small fraction of the total energy at late times.}
Similarly, the time it takes to reach shock breakout after core collapse ($\tsb$) scales with $\te$, such that 
\begin{equation}
\tsb \approx 0.91\days\ \Rfh \Efoe^{-1/2}\Mten^{1/2},
\label{eq:tsbscaling}
\end{equation}
where $\Rfh = R/500\Rsun$, $\Efoe=\Eexp/\foe$, and $\Mten=\Mej/10\Msun$, 
and the dimensionful prefactor comes from a linear fit to our numerical models. This timescale is primarily 
a property of the models, but would observationally correspond to the difference in time between the first neutrino signal 
from core collapse and the first detection in the electromagnetic spectrum from shock breakout.

Following \citet{Kasen2009}, in the limit of no accumulated heating of the ejecta due to $\Ni$ decay, 
the luminosity on the plateau (taken here to be at day 50, denoted $\Lfifty$) is set by the total internal 
energy ($\Eint$) to be radiated out divided by the duration of the plateau:
\begin{equation}
\Lfifty=\frac{\Eint(\tpt)}{\tpt}=\frac{E_0 \te}{\tpt^2},
\end{equation}
where $\tpt$ is the duration of the plateau, $E_0\approx\Eexp/2$ is the initial internal energy of the 
ejecta at shock breakout, and the second equality comes from assuming the internal energy evolution for homologous 
expansion (where $r(t)\approx v t$, for a Lagrangian fluid element with constant velocity $v$) 
in a radiation-dominated plasma, $E_0 \te = E(\tpt)\tpt$.  

Here we compare to the analytics of \citet{Popov1993},
which consider the effects of both H recombination and radiative diffusion. 
Historically, analytic scalings which ignore recombination \citep{Arnett1980} or radiative 
diffusion \citep{Woosley1988,Chugai1991} have also been considered. 
These scalings are also detailed in \citet{Kasen2009} and \citet{Sukhbold2016}.
From a 2-zone model including an optically thick region of expanding ejecta behind the photosphere
and an optically thin region outside the photosphere, Popov finds that the luminosity on the plateau (here taken at day 50)
and duration of the plateau should scale as 
\begin{equation} \label{eq:PopovScalings}
\begin{split}	
\Lfifty&\propto M^{-1/2}\Eexp^{5/6}R^{2/3}\kappa^{-1/3}T_{\rm I}^{4/3},\\ 
\tnone&\propto M^{1/2}\Eexp^{-1/6}R^{1/6}\kappa^{1/6}T_{\rm I}^{-2/3},
\end{split}
\end{equation} 
where $\kappa$ is the opacity in the optically thick component of the ejecta, and $T_{\rm I}$ is the ionization temperature of 
Hydrogen, and $M$ is the relevant mass (which could depend on the extent to which H is mixed throughout the ejecta). 
\citet{Kasen2009} recovers a similar set of scalings from their models:
\begin{equation} \label{KasenScalings}
\begin{split}
\Lfifty&\propto \Mej^{-1/2}\Eexp^{5/6}R^{2/3}\XHe^{1},\\
\tnone&\propto \Mej^{1/2}\Eexp^{-1/4}R^{1/6}\XHe^{1/2},
\end{split}
\end{equation}
where $\XHe$ is the mass fraction of He. 
There is some disagreement in the literature as to whether the mass 
$M$ used in the Popov scalings should be the mass of the hydrogen-rich 
envelope ($\Menv$) or the mass of the ejecta ($\Mej$). 
\citet{Sukhbold2016}, for example, use $\Menv$ in recreating these scalings, since recombination in the
Hydrogen-rich envelope drives the evolution of the supernova, 
with little contribution from the hydrogen-poor innermost ejecta coming from
the core. However, in our models, the relevant mass is the \emph{total} ejecta mass $\Mej$,
as we see mixing of hydrogen deep into the interior of the star and core elements 
into the envelope due to RTI. Since hydrogen recombination thus plays a significant role in setting the 
temperature throughout the entirety of the ejecta, it is the entire ejecta mass that is used in
the scalings we derive later. Additionally, we make the assumption that by day 50, 
$\Eexp$ is equal to the kinetic energy of the ejecta. 

Popov also recommended assuming that the observed photospheric velocity of the supenova ejecta should scale like $\vSN$,
such that $\vPh~\propto~(\Eexp/\Mej)^{1/2}$. However, this scaling, which does describe the \textit{typical} velocity of the SN
ejecta, should not be used when describing photospheric velocities at a fixed time, 
for reasons we discuss in Section \ref{sec:VELOCITIES}.

The above scalings do not take into account additional heating by the radioactive decay chain of $\Ni$, 
which does not significantly affect the luminosity on the plateau, but does extend the duration of 
the plateau by heating the ejecta at late times. We discuss more detailed expectations for the effects of 
$\Ni$ in Appendix \ref{sec:IIp_equations}, and its impact on our models in Section \ref{sec:DURATIONS}. 
This correction is typically written as 
\begin{equation}
\tpt = \tnone\times \frad^{1/6},
\label{eq:frad}
\end{equation}
where $\frad$ can be expressed as
\begin{equation} \label{NiCorrection1}
\frad =  1 + C_f \MNi\, \Mej^{-1/2} \Eexp^{-1/2} R^{-1},
\end{equation}
and $C_f$ is a numerical prefactor which encodes the energy and decay time of the $\Ni$ decay chain (\citealt{Kasen2009, Sukhbold2016}; 
and Appendix \ref{sec:IIp_equations}).

\begin{figure}
\centering
\includegraphics[width=\columnwidth]{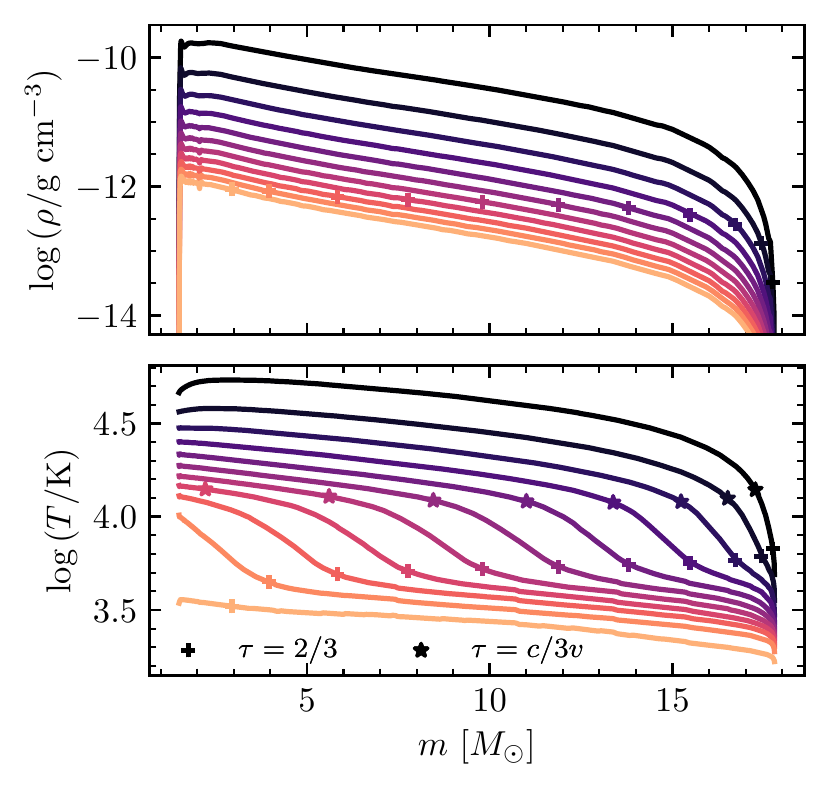} 
\caption{Density (upper panel) and temperature (lower panel) of the ejecta at every 10 days in the evolution starting at day 20
for our M16.3\_R608 model exploded with $\foe$ and $\MNi=0.03\Msun$. Darker colors indicate earlier times. Also plotted are the
photosphere ($\tau=2/3$; plus markers), and the location where $\tau=c/3v$ (star markers), which is shown only
on the temperature plot to reduce clutter, to demonstrate the extent of the region 
where the photon diffusion time is shorter than the expansion time.}
\label{fig:UtrobinPlot}
\end{figure}

These scaling relationships serve as a useful guide when modeling Type IIP supernova light curves. 
However, complexities arising from changes in temperature profiles, density profiles, realistic distributions of
important elements such as H and $\Ni$, and stellar structure can lead to differences between these simple analytic 
expectations and numerical models. For example, the Popov analytics are derived for emission from a 
two-zone model with an optically thick inner region with a single opacity $\kappa$ and an optically thin outer region
and a flat density profile. More realistic evolution of the temperature and density profiles of one of our SN ejecta models 
is shown in Figure \ref{fig:UtrobinPlot}, akin to Figure 11 of \citet{Utrobin2007}. 
Thus, in the following sections we aim to provide expressions which relate observables to the physical properties 
of the explosions, namely the progenitor $R$, $\Mej$, and $\Eexp$, while
also capturing the ejecta structure underlying these events.

\newpage
\section{Luminosity at day 50\label{sec:LUMINOSITIES}}

\begin{figure*}
\centering
\includegraphics{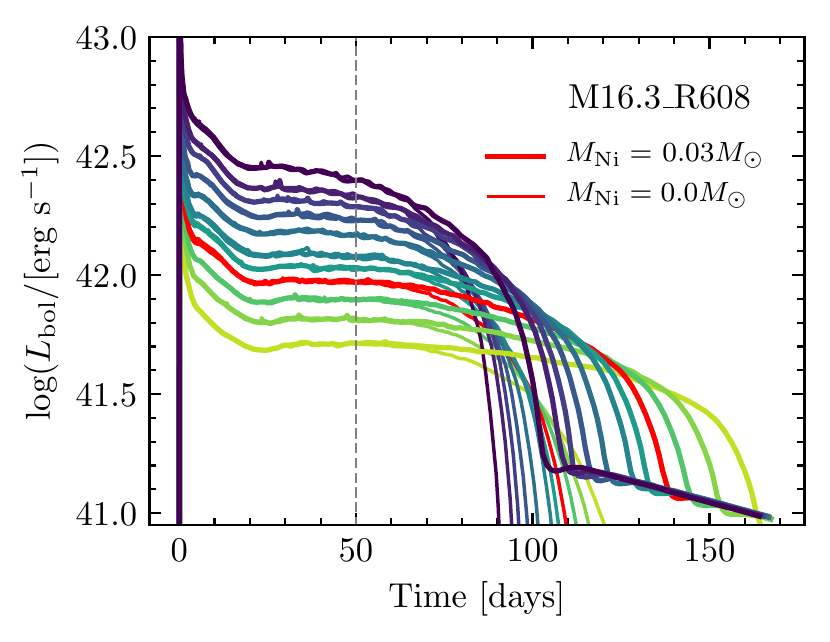} 
\includegraphics{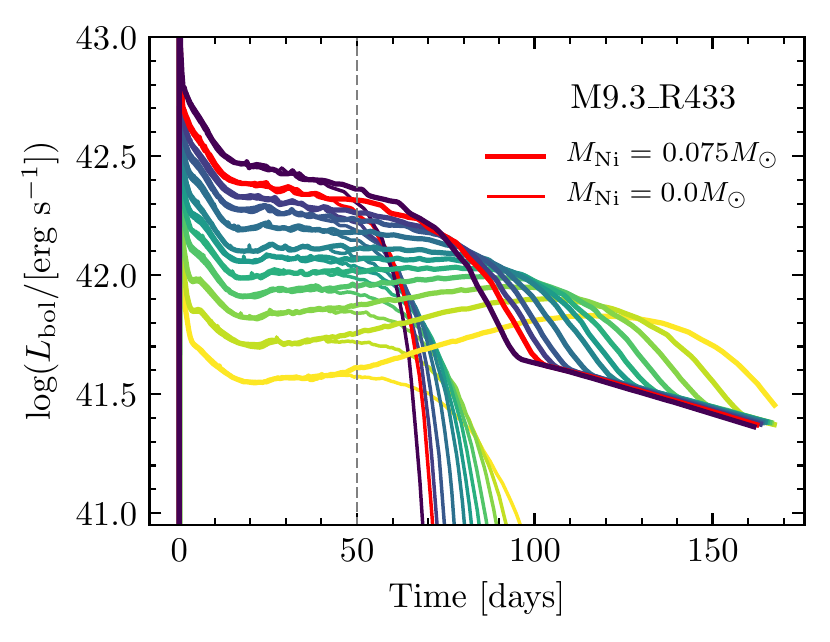} 
\caption{Light curves for increasing explosion energies of our M16.3\_R608 model with 
$\MNi=0.03\Msun$ (left) and our smallest model, M9.3\_R433, with $\MNi=0.075\Msun$ (right).
Thin lines correspond to the same explosions, but with no $\Ni$.
The red lines correspond to the models with the typical $\MNi$ given 
$\Lfifty$, from the relationship in \citet{Pejcha2015b}, and their no-nickel counterparts.}
\label{fig:Nickellight curves}
\end{figure*}

We use the bolometric luminosity 50 days after shock breakout, $\Lfifty$, as 
our diagnostic for the plateau luminosity, as in most cases, this is beyond the 
time where shock heating of CSM would affect the 
luminosity \citep{Morozova2017}. 
Moreover, for all but one progenitor model, increasing the amount of $\Ni$ has a negligible 
impact on $\Lfifty$, as the internal energy at day 50 of the outer region of the ejecta is still
dominated by the initial shock. However, in explosions where the plateau is 
naturally short, there can be marginal, but noticeable, additional luminosity at day 50 from 
$\Ni$ decay. This can be seen in Figure \ref{fig:Nickellight curves},
which shows the differences between a selection of light curves and the corresponding light 
curves with no $\Ni$. We show light curves for M16.3\_R608, a typical model with a 
typical $\MNi=0.03\Msun$ (left), and for high $\MNi=0.075\Msun$ explosions of the only progenitor 
model in which we see significant deviation in $\Lfifty$ as a result of $\Ni$ heating, 
M9.3\_R433 (right), where $\Lfifty$ varies by up to 15\% between an explosion with no 
$\Ni$ and one with $\MNi=0.075\Msun$. Noting this, we choose a moderate, constant value 
of $\MNi =0.03\Msun$ typical of observed events \citep{Muller2017}, and calculate how 
$\Lfifty$ scales with $\Mej$, $\Eexp$, and $R$. 

\begin{figure}
\centering
\includegraphics[width=\columnwidth]{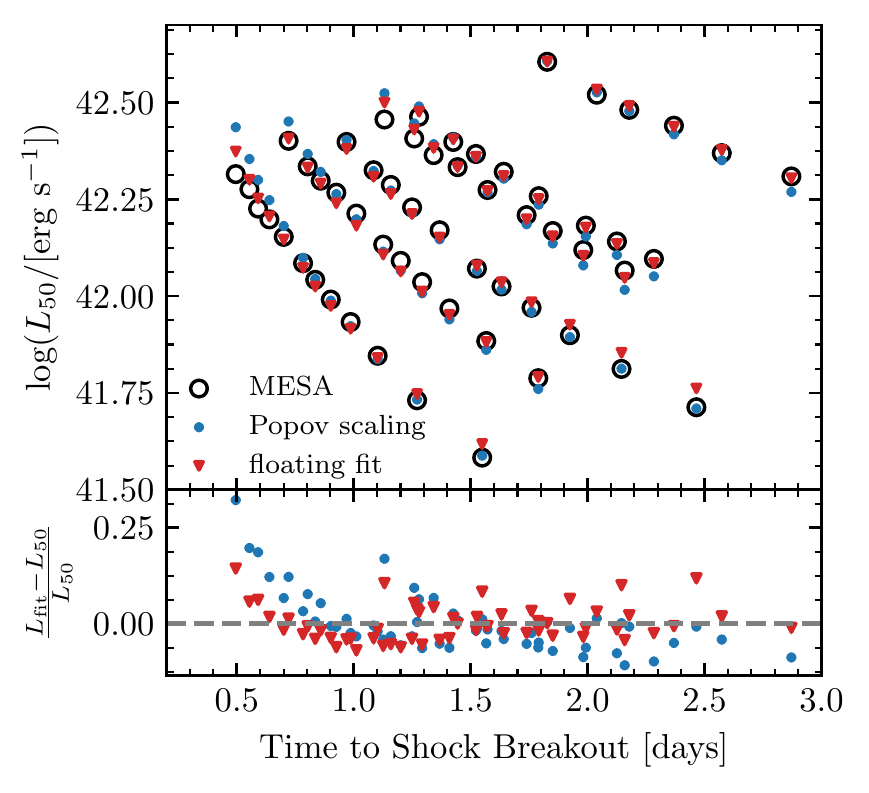} 
\caption{Fitting formulae predictions for $L_{50}$ (colored markers) compared with the model 
$\Lfifty$ (black circles; upper panel), and their residuals (lower panel) for 
our suite of 57 explosions at constant $\MNi=0.03\Msun$. Each diagonal family of points 
reflects one progenitor model exploded with different energies.}
\label{fig:L50fits}
\end{figure}

We fit two formulae to our sample suite of 57 explosions. First, we assume the power law 
coefficients of \citet{Popov1993}, and let the normalization float, finding
\begin{align}
\begin{split}
&\log(\Lfifty/[{\rm erg\, s^{-1}}])=\\
&42.18 - \frac{1}{2}\log \Mten + \frac{5}{6}\log \Efoe + \frac{2}{3}\log \Rfh,
\end{split}
\label{eq:LPopovFit}
\end{align} 
where $42.18$ is a linear fit from our models and logarithms are base 10, with
$\Mten=\Mej/10 \Msun$, $\Rfh=R/500 \Rsun$, and $\Efoe = \Eexp/10^{51}$\,ergs.
For these models, root mean square (RMS) deviations of $\Lfifty$ from values derived by applying Equation 
\eqref{eq:LPopovFit}, corresponding to the blue points in Figure \ref{fig:L50fits}, are 
$7.9\%$, with a maximum deviation in $\Lfifty$ of $32\%$. 
The normalization for an explosion with $\Mten = \Rfh = \Efoe = 1$ is comparable to
but somewhat lower than the value of 42.27 given in \citet{Sukhbold2016} (who use $\Menv$ 
rather than $\Mej$), 
as well as the value of 42.21 calculated in \citet{Popov1993} for default H recombination 
temperatures and opacities. \citet{Kasen2009} give a value of 42.10$+\log(X_{\rm He}/0.33)$, 
letting $X_{\rm He}$ range from 0.33 to 0.54. 

We perform a second fit for the normalization \textit{and} the power laws in $\Mten,\ \Efoe,$ 
and $\Rfh$, and recover scalings that are similar to those in Equation 
\eqref{eq:LPopovFit}. We find a slightly shallower scaling with $\Mej$ and $\Eexp$, 
and a slightly steeper dependence on $\Rfh$:
\begin{align} 
\begin{split}
&\log(\Lfifty/[{\rm erg\, s^{-1}}]) = \\ 
&42.16 - 0.40 \log M_{10} + 0.74 \log E_{51} + 0.76 \log R_{500}, 
\end{split}
\label{eq:LFloatingFit}
\end{align}
where the normalization and power law coefficients are fit from our models. 
The  RMS deviation of the models from Equation \eqref{eq:LFloatingFit}, shown as red triangles 
in Figure \ref{fig:L50fits}, is 4.7\%, with deviations not exceeding 14.3\% for any model 
with $\MNi=0.03\Msun$. This is a better fit than the one that assumes the exact Popov scaling. 

The luminosities of our models, as they compare to the fitted formulae, 
are shown in Figure \ref{fig:L50fits}. Most models agree with the 
Popov scaling, while the Popov scaling overpredicts $\Lfifty$ in low-ejecta mass 
high-explosion energy cases. The x-axis of Figure \ref{fig:L50fits} is $\tSB$, 
chosen because it scales with explosion energy for a fixed ejecta mass 
and radius (Equation \eqref{eq:tsbscaling}), and increases with increasing 
$\Mej$ and $R$, visually distinguishing the six different progenitor 
models and different explosion energies. 

Although the presence of $\Ni$ does not affect light curve properties at day 50 in a majority of 
models, in a few explosions there is slight variation in $\Lfifty$ introduced by the 
extra heating from $\Ni$ (seen in Figure \ref{fig:Nickellight curves}). Because this effect is only 
distinctly noticeable in our model with the smallest values of $\Mej$ and $R$, this can lead to variations 
in our recovered power laws when fitting to different fixed $\Ni$ masses. However, this correction is typically 
small, falling within the scatter in which our models agree with the fitted formulae. We find that the 
power law scalings of Equation \eqref{eq:LFloatingFit} describe all models with $\MNi$ ranging from 
$0.0-0.075\Msun$ within 18.2\%, with  RMS deviations of 4.8\%, where the largest deviations occur
in events where $\MNi=0.0\Msun$, which are not consistent with any observed Type IIP SNe. 

We now use Equation \eqref{eq:LFloatingFit} to compare 
our \MESA+\STELLA\ results to models from other software instruments. 
In Table \ref{tbl:AllLuminosityTables}, we show our predictions for $\Lfifty$ compared against luminosities from the \MESA+\CMFGEN\ models
(without Duffell RTI) of \citet{Dessart2013}, the \Kepler+\Sedona\ models of \citet{Kasen2009}, and the 
\MESA+\CMFGEN\ models in \citet{Lisakov2017}. In general, the disagreement between our 
formula and these other models is similar to the scatter within our own models, 
with the exception of the two lowest-energy explosions in \citet{Kasen2009} and the low 
luminosity suite in \citet{Lisakov2017}. Equation \eqref{eq:LFloatingFit} agrees with
the \citet{Dessart2013} models with an  RMS error of 9\%, but slightly underpredicts
luminosity in a majority of cases. Compared to the models of \citet{Kasen2009}, 
Equation \eqref{eq:LFloatingFit} gives  RMS errors of 17\%  with no clear under- or overprediction. 
The low-luminosity models from \citet{Lisakov2017} have greater disagreement, with  RMS errors 
23\% from Equation \eqref{eq:LFloatingFit}. This is not surprising, as on the low-luminosity end, 
our treatment of fallback discussed in Section~\ref{sec:MODELS} and Appendix~\ref{sec:FALLBACK} 
excludes most models in this region of parameter space from our fitting formulae, as significant fallback 
after the initial core collapse is often seen for low explosion energies.

\begin{table*}
\caption{Comparison of Equation \eqref{eq:LFloatingFit} to \CMFGEN\ models 
from \citet{Dessart2013}, \Kepler+\Sedona\ models from \citet{Kasen2009}, and 
low-luminosity \CMFGEN\ models from \citet{Lisakov2017}.  Bolometric luminosities at day 
50 for \citet{Dessart2013} are recovered from light curves provided by L. Dessart (private communication). 
These luminosities are compared to Equation \eqref{eq:LFloatingFit} applied to $\Mej$, $\Eexp$, 
and $R$ from the various models.}
\centering
\singlespace
\begin{tabular}{clcccccc}  
\hline\hline
Source & Model & $\Mej$ & $\Eexp$  & $R$ &  $\Lfifty$ & Equation \eqref{eq:LFloatingFit} & \% diff\\ 
& & [M$_\odot$] & [$10^{51}$ erg] & [$R_\odot$] & [10$^{42}$ erg s$^{-1}$] & [10$^{42}$ erg s$^{-1}$] &   \\ 
\hline
Dessart+13  & m15Mdot    & 10.01    & 1.28   & 776     & 2.55    & 2.40 & -5\%   \\ 
			& m15 	   & 12.48    & 1.27   & 768     & 2.56      & 2.17 & -15\%   \\ 
			& m15e0p6    & 12.46    & 0.63   & 768     & 1.19    & 1.29 & 8\%   \\ 
			& m15mlt1    & 12.57   & 1.24   & 1107     & 3.13     & 2.81 & -10\%  \\ 
			& m15mlt3    & 12.52   & 1.34   & 501      & 1.61      & 1.63 & 1\%  \\ 
			& m15os  	   & 10.28   & 1.40   & 984      & 3.49     & 3.05 & -12\%  \\ 
			& m15r1  	   & 11.73   & 1.35   & 815      & 2.62     & 2.44 & -7\%   \\ 
			& m15r2      & 10.39   & 1.34   & 953      & 3.30      & 2.87 & -13\%  \\  
			& m15z2m3    & 13.29   & 1.35   & 524      & 1.70     & 1.65 & -3\%  \\  
			& m15z4m2    & 11.12   & 1.24   & 804      & 2.48     & 2.31 & -6\%  \\  
			& s15N       & 10.93   & 1.20   & 810      & 2.51     & 2.29 & -9\%   \\  
			& s150       & 13.93   & 1.20   & 610      & 2.47   & 2.29 &  -8\%  \\  

\hline
K\&W 2009 &M12\_E1.2\_Z1    & 9.53    & 1.21   & 625     & 1.91    & 1.99 & 4\%    \\ 
          &M12\_E2.4\_Z1    & 9.53    & 2.42   & 625     & 3.67    & 3.33 & -9\%   \\ 
          &M15\_E1.2\_Z1    & 11.29   & 1.21   & 812     & 2.16    & 2.27 & 5\%    \\ 
          &M15\_E2.4\_Z1    & 11.29   & 2.42   & 812     & 4.35    & 3.80 & -12\%  \\ 
          &M15\_E0.6\_Z1    & 11.25   & 0.66   & 812     & 1.26    & 1.45 & 15\%   \\ 
          &M15\_E4.8\_Z1    & 10.78   & 4.95   & 812     & 7.80    & 6.59 & -15\%  \\ 
          &M15\_E0.3\_Z1    & 11.27   & 0.33   & 812     & 0.59    & 0.87 & 46\%   \\ 
          &M20\_E1.2\_Z1    & 14.36   & 1.22   & 1044    & 2.61    & 2.52 & -4\%   \\ 
          &M20\_E2.4\_Z1    & 14.37   & 2.42   & 1044    & 4.85    & 4.18 & -13\%  \\ 
          &M20\_E0.6\_Z1    & 14.36   & 0.68   & 1044    & 1.40    & 1.63 & 17\%   \\ 
          &M20\_E4.8\_Z1    & 14.37   & 4.99   & 1044    & 8.57    & 7.16 & -17\%  \\ 
          &M25\_E1.2\_Z0.1  & 13.27   & 1.26   & 632     & 1.67    & 1.82 & 8\%    \\ 
          &M25\_E2.4\_Z0.1  & 13.24   & 2.48   & 632     & 3.08    & 3.00 & -2\%   \\ 
          &M25\_E0.6\_Z0.1  & 13.28   & 0.65   & 632     & 0.86    & 1.11 & 29\%   \\ 
          &M25\_E4.8\_Z0.1  & 13.18   & 4.90   & 632     & 5.31    & 4.98 & -6\%   \\ 

\hline
Lisakov+17 &X     & 8.29   & 0.25   & 502    & 0.446    & 0.550 & 24\%   \\ 
           &XR1   & 8.08   & 0.26   & 581    & 0.513    & 0.643 & 23\%   \\ 
           &XR2   & 7.90   & 0.27   & 661    & 0.592    & 0.737 & 24\%   \\ 
           &XM    & 9.26   & 0.27   & 510    & 0.423    & 0.567 & 34\%   \\ 
           &YN1   & 9.45   & 0.25   & 405    & 0.381    & 0.446 & 17\%   \\ 
           &YN2   & 9.45   & 0.25   & 405    & 0.381    & 0.446 & 17\%   \\ 
           &YN3   & 9.45   & 0.25   & 405    & 0.375    & 0.446 & 19\%   \\\hline  
\end{tabular}
\doublespace
\label{tbl:Lisakov2017}
\label{tbl:AllLuminosityTables}
\end{table*}

\section{Plateau Duration and $\ET$ \label{sec:DURATIONS}}
Although the plateau duration $\tpt$ is theoretically motivated by \citet{Popov1993, Kasen2009}, 
and others, it is important to reliably extract it from our models as well 
as observations. We discuss two ways of extracting $\tpt$, one defined by observables, 
and the other extracted from properties of the theoretical 
models, which we use to evaluate the impact of $\Ni$.

\begin{figure}
\centering
\includegraphics[width=\columnwidth]{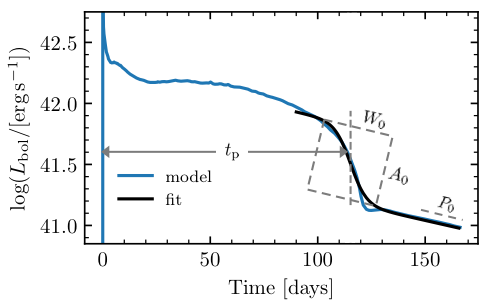} 
\caption{Diagram of fitting Equation \eqref{eq:FermiFunction} to find the 115 day duration of the plateau 
for our model M16.3\_R608 exploded with $\Eexp=\foe$ and $\MNi=0.03\Msun$. 
Fitting parameters are labelled, but we only ascribe physical significance to $t_{\rm p}$.} 
\label{fig:ObserverPlateau}
\end{figure}

For a definition which can be applied to observed or calculated light curves, we follow \citet{Valenti2016}, 
fitting the following functional form to the logarithm, $y(t)$, of the bolometric luminosity around the
fall from the plateau:
\begin{equation}
y(t) = \log_{10}(\Lbol)= \frac{-A_0}{1 + e^{(t - \tpt)/W_0}} + (P_0 \times t) + M_0.
\label{eq:FermiFunction}
\end{equation}
An example is shown in Figure \ref{fig:ObserverPlateau}. We use the python 
routine \texttt{scipy.optimize.curve\_fit} to fit the light curve starting at the time when the 
luminosity evolution is 75\% of the way to its steepest descent, defined when $\dif\log{\Lbol}/\dif t$ 
is at its most negative after the initial drop at shock breakout, which occurs 
shortly before transitioning to the nickel tail. We fix the value of $P_0$ to be the 
slope on the $\Ni$ tail. We interpret the fitting parameter $\tpt$ to be 
the plateau duration. 

\begin{figure}
\centering
\includegraphics[width=\columnwidth]{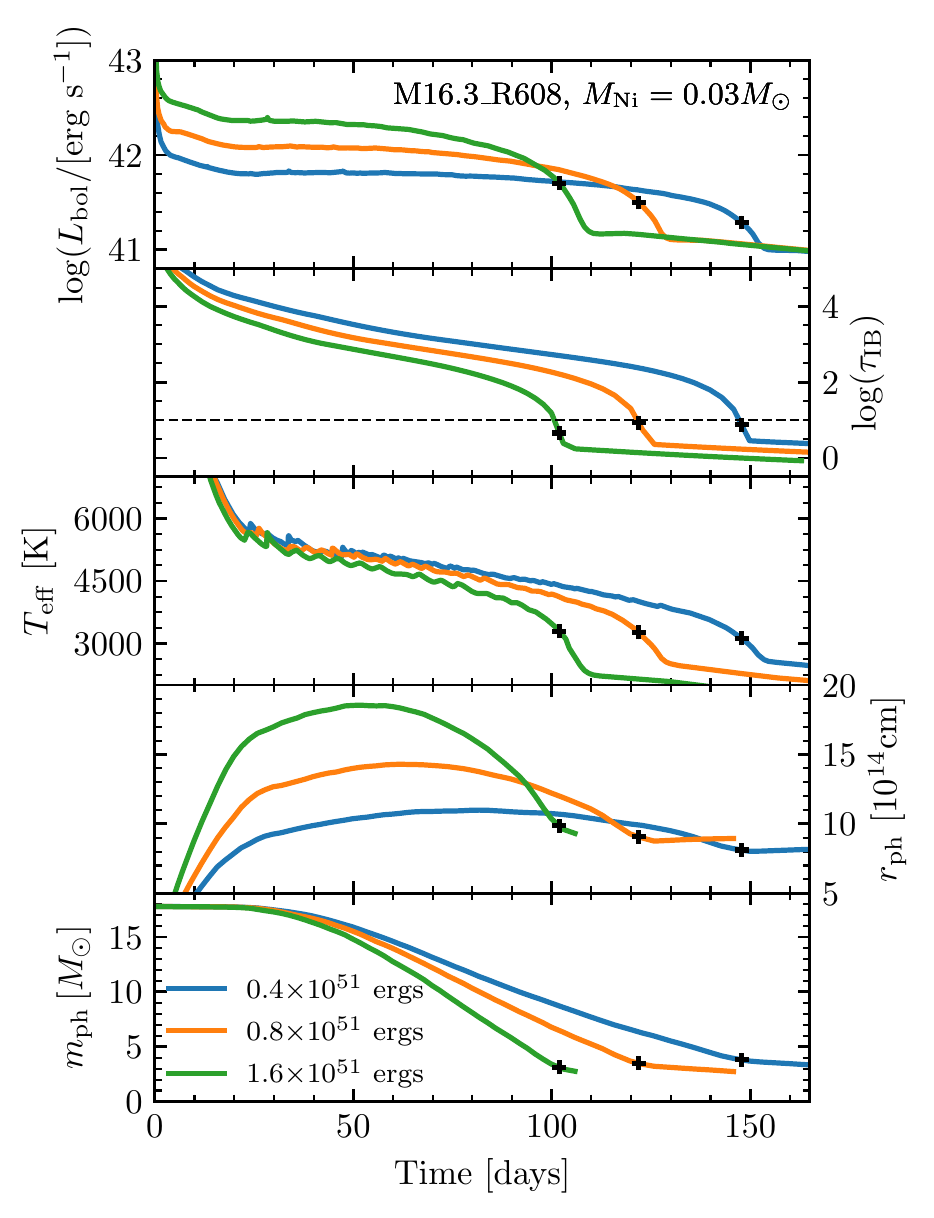} 
\caption{Evolution of luminosity, optical depth at the inner boundary ($\tau_{\rm IB}$), effective 
Temperature ($\Teff$), photospheric radius ($r_{\rm ph}$), and mass coordinate of the photosphere 
($m_{\rm ph}$) for our M16.3\_R608 model, exploded 
with $\MNi=0.03\Msun$ at three different explosion energies, showing the transition to the end 
of the plateau. Black plus markers indicate the end 
of the plateau determined from fitting Equation \eqref{eq:FermiFunction}. The thin dashed line in 
the second panel indicates $\tau_{\rm IB}=10$.}
\label{fig:ttaudefinition}
\end{figure}

As the recombination-powered photosphere moves into the innermost ejecta, the optical 
depth at the inner boundary declines orders of magnitude and the photospheric temperature 
plummets. This transition, shown in Figure \ref{fig:ttaudefinition}, is the physical end of the 
plateau. Thus, for our modeling definition of the plateau duration, we 
use the time, post-shock breakout, when the optical depth $\tauIB$ through the ejecta becomes $< 10$. 
This time will be denoted hereafter as $\ttau$, and can be used as a metric for plateau duration
when comparing to models where there is no $\Ni$, where Equation \eqref{eq:FermiFunction} does not 
accurately capture the fall from the plateau. As shown by the black markers 
in Figure \ref{fig:ttaudefinition}, the observable $\tpt$ roughly corresponds to 
the physical end of the plateau phase around $\ttau$. 
Across all progenitor models, explosion energies, and nonzero nickel masses which we consider, 
RMS differences between $\ttau$ and $\tpt$ are 4.1\% and all differences are within $\pm 7\%$.
 
\subsection{Impact of $\Ni$ on plateau duration in our models}
\begin{figure}
\centering
\includegraphics[width=\columnwidth]{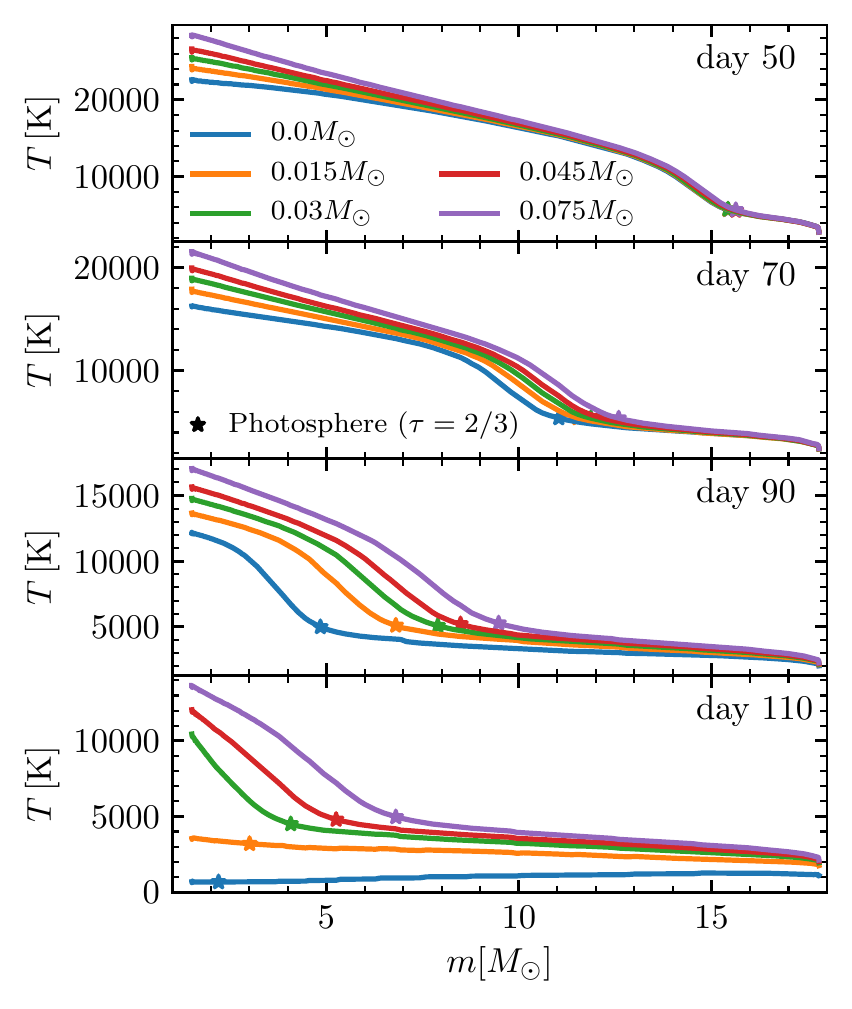} 
\caption{Ejecta temperature profiles at different times with increasing amounts of $\Ni$ for
our M16.3\_R608 model exploded with $10^{51}$ ergs. 
The location of the photosphere is shown for each model by the colored stars.}
\label{fig:TTimeLapse}
\end{figure}

The presence of radioactive $\Ni$ prolongs the photospheric evolution and extends 
the plateau by providing extra heat to the ejecta. This is shown in Figure 
\ref{fig:TTimeLapse}, 
where we show ejecta temperature profiles of the same SN explosion with different $\MNi$. 
At day 50, the photosphere for all models remains in the outer ejecta, 
where there is very little $\Ni$. At later times, the photosphere has moved in farther 
for models with lower $\MNi$, whereas additional heat from the $\Ni$ decay chain causes the 
recombination-powered photosphere to move in more slowly in models with higher $\MNi$. 

\begin{figure}
\centering
\includegraphics[width=\columnwidth]{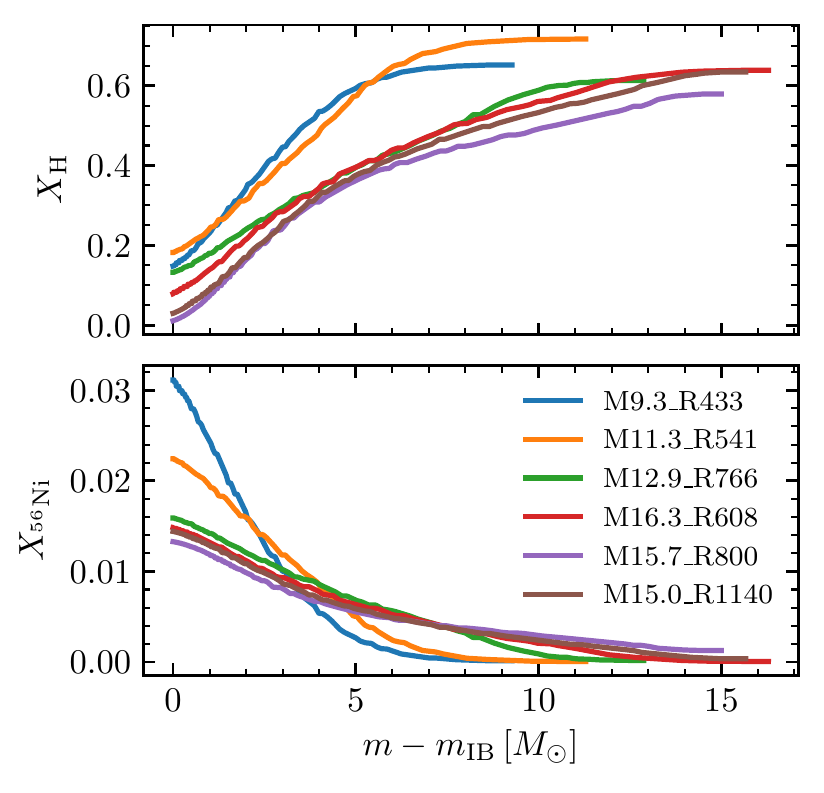} 
\caption{The diversity of H (upper panel) and 
$\Ni$ (lower panel) profiles for our standard suite of models, exploded 
with our highest $\Eexp$ and $\MNi$. The x-axis is the mass coordinate 
within the ejecta.}
\label{fig:HandNiProfiles}
\end{figure}

The analytics in Appendix \ref{sec:IIp_equations} and Section \ref{sec:PEDAGOGY} 
treat the ejecta as a single zone, with heating from 
$\Ni$ decay throughout. However, $\Ni$ is more highly concentrated in the center of the ejecta.
Thus, heat from $\Ni$ decay remains trapped in the optically thick inner region, 
extending the plateau more at late times. This more concentrated $\Ni$ 
heating should have a more significant impact on the plateau duration than it would for an analytic 
one-zone model, as the internal energy of the inner ejecta is more relevant than that of the 
ejecta as a whole at the end of the plateau. 
Figure \ref{fig:HandNiProfiles} shows the diversity of asymptotic $\Ni$ 
and Hydrogen distributions within our standard suite of models at handoff to \STELLA\
for the highest-energy ($\Etot=2\times\foe$) highest-Nickel ($\MNi=0.075\Msun$) cases.

\begin{figure}
\centering
\includegraphics[width=\columnwidth]{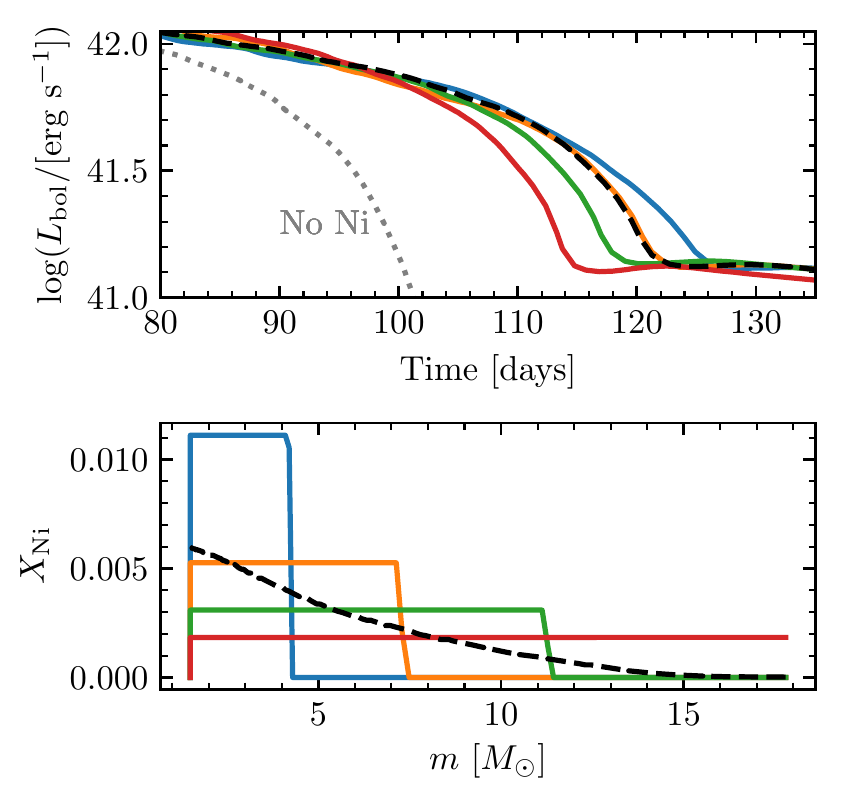} 
\caption{Light curves at the end of the plateau (upper panel) and $\Ni$ 
distributions (lower panel) for the M16.3\_R608 model exploded with $\Eexp=\foe$ 
and $\MNi=0.03\Msun$, for the fiducial explosion (black dashed line), compared 
to models where $\Ni$ is re-distributed by hand out to some fraction of the 
ejecta at the time of shock breakout (solid colored lines). The light curve 
for the same explosion with no Ni is given by the grey dashed line in the 
upper panel.}
\label{fig:TheoristDelight}
\end{figure}

Moreover, the distribution of $\Ni$, which can vary amongst different progenitors depending on
core structure and mixing, can also introduce inherent scatter to the plateau duration
\citep{Kozyreva2018}. Figure \ref{fig:TheoristDelight} shows light curves and $\Ni$ profiles for
the M16.3\_R608 model exploded with $\Eexp=\foe$ and $\MNi=0.03\Msun$, where the same $\Ni$ mass is 
re-distributed by hand at the time of shock breakout out to some fraction of the ejecta. 
Although this exercise spans a greater diversity in $\Ni$ concentration than any of our models, 
we see for this otherwise unexceptional light curve that the plateau duration $\tpt$ can vary by 
almost 10 days.

A full examination of the effects of changing the 
distributions in the framework of the Duffell RTI prescription 
\citep{Paxton2018} is beyond the scope of this paper, 
and will be the subject of future study (P. Duffell et al. in Prep.). 
Here we examine the impact of $\Ni$ on the value of $f_{\rm rad}$ in
$\tpt = t_0 \times f_{\rm rad}^{1/6}$ (Equation \eqref{eq:frad}), 
where $t_0$ is the plateau duration for the same explosion with no $\Ni$. 

Following \citet{Kasen2009}, \citet{Sukhbold2016}, and others, $\Ni$ extends the plateau as 
\begin{equation}
\tpt/t_0 = (1 + C_f \MNi \Efoe^{-1/2} \Mten^{-1/2} \Rfh^{-1})^{1/6},
\label{eq:KasenScaling}
\end{equation}
we can extract $C_{f}$ by fitting to our models using the $\ttau$ definition of plateau duration. 
We consider all six progenitor models with explosion energies sufficient to 
cause minimal fallback, with $\MNi/\Msun$=0.0, 0.015, 0.03, 0.045, 0.06, and 0.075. 
We exclude models where the plateau is so long that the Nickel tail does not 
appear at any point in our simulations, and models which have a less than half 
a decade drop in $\Lbol$ from day 50 to the top of the nickel tail, as no 
such events have been observed.\footnote{This primarily excludes models at
high nickel masses and low explosion energies, specifically:
M9.3\_R433: $\Efoe=0.2$ with $\MNi=0.075\Msun$;
M11.3\_R541: $\Efoe=0.3$ with $\MNi=0.06\Msun$ and $\MNi=0.075\Msun$; and
M16.3\_R608: $\Efoe=0.3$ with $\MNi=0.045\Msun$, $\MNi=0.06\Msun$, and $\MNi=0.075\Msun$; 
$\Efoe=0.4$ with $\MNi=0.06\Msun$ and $\MNi=0.075\Msun$;
and $\Efoe=0.4$ with $\MNi=0.075\Msun$.}

\begin{figure}
\centering
\includegraphics[width=\columnwidth]{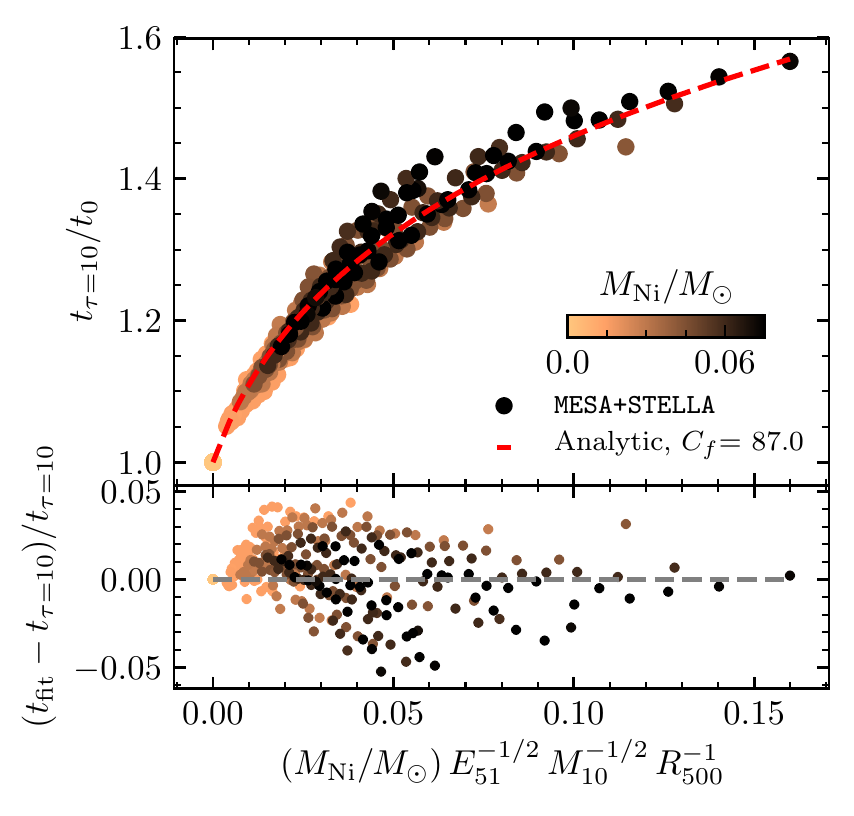} 
\caption{Plateau duration when the optical depth at the inner boundary $\tauIB=10$, 
divided by plateau duration for the same explosion with no $\Ni$, as compared to the analytic 
scaling Equation \eqref{eq:KasenScaling} (red dashed line) with $C_f$ determined from our fits. Color 
corresponds to $\MNi$ in units of solar masses. Deviations of each of the models from this 
relationship are shown on the lower panel.}
\label{fig:KasenNickel}
\end{figure}

This gives a total of 332 light curves including the 57 with no $\Ni$, which we 
compare to the light curves of identical explosions with no $\Ni$. 
Figure \ref{fig:KasenNickel} shows the ratio of the plateau duration, 
$\ttau$, of each of these light curves compared to $\ttau$ for an identical 
explosion with no $\Ni$, $t_0$, following \citet{Kasen2009} but with our suite 
of 332 model light curves. We recover $C_f \approx 87$, which is an order of 
magnitude larger than the approximate lower bound $C_f \approx 7.0$ derived in 
Appendix \ref{sec:IIp_equations}, and roughly a factor of 4 larger than 
$C_f=24$ (derived in \citealt{Kasen2009}, typographical error corrected in 
\citealt{Sukhbold2016}). 
This likely results from the different $\Ni$ mass distributions 
in our models from those in \citet{Kasen2009}. As demonstrated in Figure 
\ref{fig:TheoristDelight}, this can yield significant differences in the 
plateau duration. Our fit shows similar scatter for all $\Ni$ masses 
considered, with more scatter introduced by intrinsic differences among the 
individual models than by the changing $\MNi$. For $\MNi \gtrsim 0.03\Msun$, 
this typically leads to a 20 - 60\% increase in the plateau duration. 

\subsection{Plateau Durations for Nickel Rich Events}

For Nickel-rich events, the $\Ni$ and $\Co$ decay dominates the internal energy of the inner 
ejecta, such that $C_f \MNi \Mten^{-1/2} \Efoe^{-1/2} R^{-1} > 1$. Assuming that $t_0$ scales 
as in \citet{Popov1993}, we can approximate
\begin{align}
\begin{split}
&\tpt \propto \Mej^{1/2} \Eexp^{-1/6} R^{1/6} \times 
(1 + C_f \MNi \Mej^{-1/2} \Eexp^{-1/2} R^{-1})^{1/6}\\
 &\rightarrow \tpt \appropto \MNi^{1/6} \Mej^{5/12} \Eexp^{-1/4}.
\label{eq:predictedtpt}
\end{split}
\end{align}

\begin{figure}
\centering
\includegraphics[width=\columnwidth]{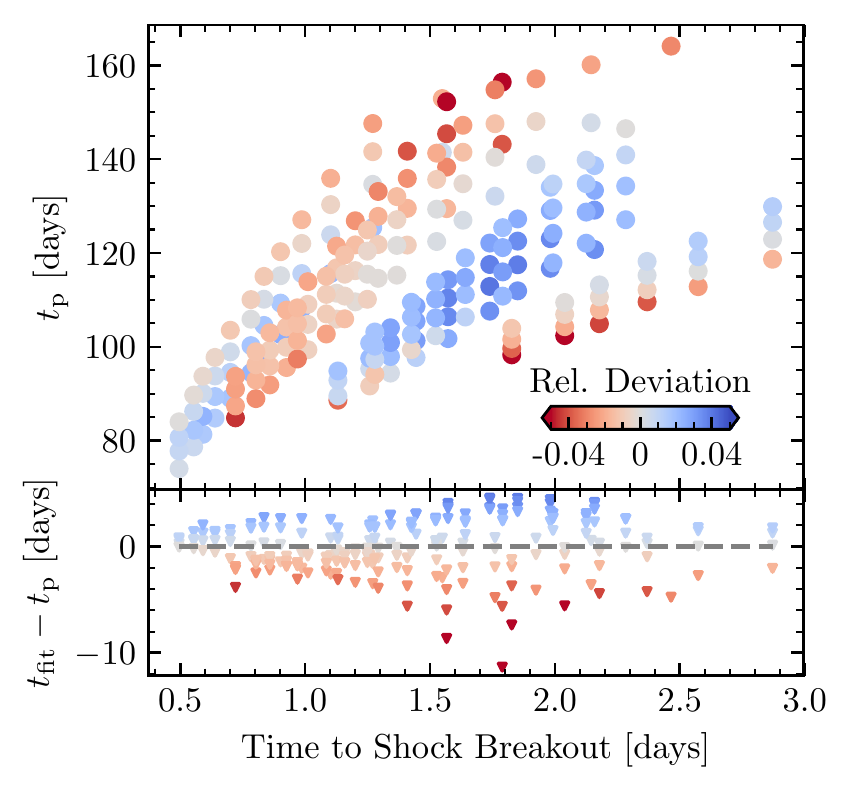} 
\caption{Plateau durations for our 218 SN light curves with $\MNi\gtrsim0.03\Msun$ (upper panel) and the
difference in plateau duration between the model $\tpt$ and the plateau duration $\tfit$ 
extracted by applying Equation \eqref{eq:tptscaling} to the $\Mej$, $\Eexp$, and $\MNi$ of each 
model (lower panel), colored by $(\tfit - \tpt)/\tpt$, the fractional deviation between $\tpt$ from the models and $\tfit$. 
The x-axis is the time the model takes to reach shock breakout.}
\label{fig:UglyPlotLessUgly}
\end{figure}

The two features of interest are the power law behavior and the disappearing scaling with the progenitor radius. 
We thus expect that the plateau duration for $\Ni$-rich events does not depend on the progenitor radius. 
To check, we perform a power law fit for $\tpt$ 
as a function of $\MNi$, $\Mej$, $\Eexp$, and $R$ for 218 model light curves 
where $\MNi\gtrsim0.03\Msun$. 
We find that $\log(\tpt/{\rm days})= 2.184 + 0.134\log(\MNi)+0.429\log(\Mten)-0.280\log(\Efoe)-
0.018\log(\Rfh)$ with  RMS deviations of 2.10\% and a maximum deviation of 8.1\%. 
Since the dynamic range in $R$ is a factor of two and the scaling is negligible, 
we perform a fit for these same models to only $\MNi$, $\Mej$, and $\Eexp$,
recovering
\begin{align}
\begin{split}
\log(\tpt/{\rm days}) = &2.184 + 0.134\log(\MNi)\\& +  0.411 \log(\Mten) - 0.282\log(\Efoe).
\label{eq:tptscaling}
\end{split}
\end{align}
These coefficients are excellent matches to the power laws in Equation 
\eqref{eq:predictedtpt}.
Our models, and their agreement with this fit, are shown in Figure 
\ref{fig:UglyPlotLessUgly}.  RMS deviations between this fit and our models are 2.13\%, 
with maximum deviation of 7.5\%. 
Typical differences between the plateau durations recovered from the fit and those 
extracted from our models are $2-5$ days, with the largest discrepancy being 11 
days, which is for a relatively low-luminosity SN with a plateau duration of 156 days. 
For all of the scaling equations of this section, the scatter in our models does 
not require that we report the fits to three decimal places; this is done for the sake of
completeness.

We also checked the agreement of Equation \eqref{eq:tptscaling} with the publicly available light curves 
from \citet{Dessart2013}. For all of those models where the light curve has a clear end of plateau and 
nickel tail, we found that our fitting formula recovers a plateau which is 7 - 20\% shorter when using 
the values for $\MNi$, $\Mej$, and $\Eexp$ reported in \cite{Dessart2013}. This amounts to a difference 
of 8 - 27 days, with the worst agreement in the case of the low-metallicity (1/10 solar) model m15z2m3, 
and the best agreement in the case of their ``new'' s15N model. The RMS difference 
in $|\tpt - \tfit|$ is 18 days, about 14\% relative to the average plateau in their models.

\subsection{Constraining Explosion Parameters with $\ET$}
\begin{figure}
\centering
\includegraphics[width=\columnwidth]{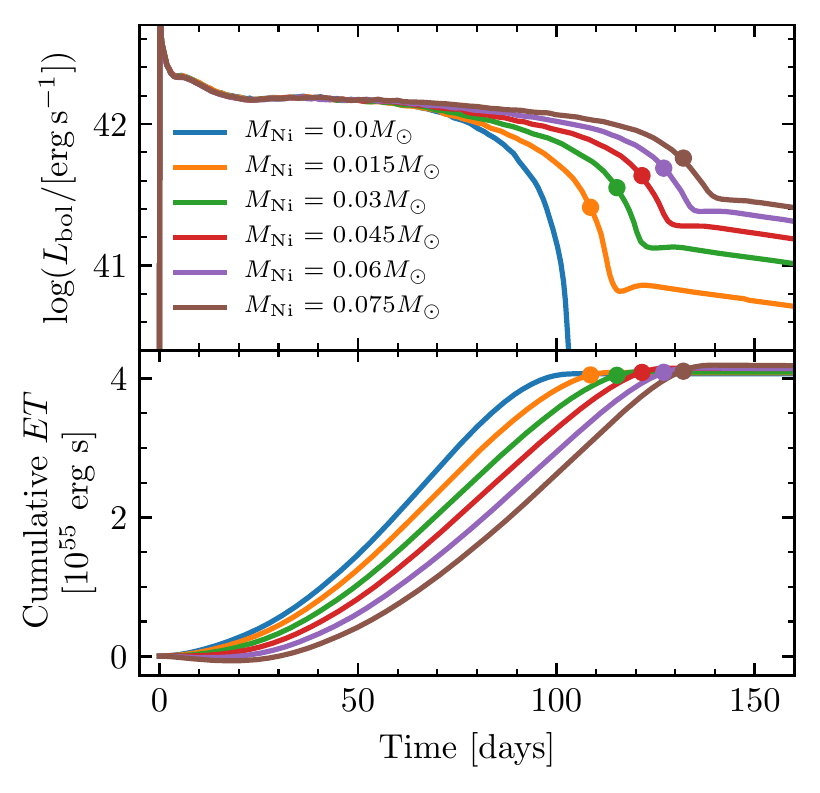} 
\caption{Impact of $\Ni$ on $\Lbol$ (upper panel) and cumulative $\ET$ integrated from shock breakout 
to the time on the x-axis (lower panel) for  M16.3\_R608 model exploded with $\Eexp=\foe$ and 
$\MNi/\Msun=0.0,\ 0.015,\ 0.03,\ 0.045,\ 0.06,$ and 0.075. Points indicate $\tpt$ 
for the events where $\MNi>0.0\Msun$.}
\label{fig:ETforDiffNic}
\end{figure}

Following the work of \citet{Shussman2016}, \citet{Nakar2016}, \citet{Kozyreva2018}, and others,
we can also express the impact of $\Ni$ on $\tpt$ in terms of the ratio of 
the time-weighted energy contribution of the $\Ni$ decay chain to the 
observable quantity $\ET$. This ratio is defined in \citet{Nakar2016} as 
\begin{equation}
\etaNi=\frac{\int_{0}^{\tpt} t \QNi(t) \dif t}{\ET},
\end{equation}
where 
\begin{equation}
\ET=\int_{0}^{\infty} t (\Lbol(t) - \QNi(t))\,\dif t,
\label{eq:ETeq}
\end{equation}
is the time-weighted energy radiated away which was generated by the initial shock and not by $\Ni$ decay, and
\begin{equation}
\QNi=\frac{\MNi}{\Msun}\left(6.45e^{-t/8.8\days} + 1.45e^{-t/113\days}\right)\times10^{43}\ {\rm erg\ s^{-1}},
\label{eq:QNi}
\end{equation}
is the instantaneous heating rate of the ejecta due to the decay chain of radioactive $\Ni$ assuming 
complete trapping given in \citet{Nadyozhin1994}, and $t$ is the time in days since the explosion. 
It is generally assumed that $\Lbol(t) = \QNi(t)$ after the photospheric phase, on the Nickel tail, 
and so the integral for $\ET$ is often expressed to be bounded at $\tNi\approx\tpt$.
We find this to be valid; see the lower panel of Figure \ref{fig:ETforDiffNic}.

Figure \ref{fig:ETforDiffNic} shows the impact of $\Ni$ on light curves and the integrated $\ET$ 
for the M16.3\_R608 model exploded with $\Eexp=\foe$ at different $\MNi$. The lower panel 
gives the cumulative $\ET$, integrated from shock breakout to the time on the x-axis. 
Most of the contribution to $\ET$ comes from luminosity on the plateau, 
with little contribution at early times ($t<20\,\days$) and no contribution from the Nickel tail. 
In the very $\Ni$-rich case, the cumulative integral may dip slightly negative around day 20, 
as radiative cooling is briefly less efficient than heating from the $\Ni$ decay chain 
($\Lbol<\QNi$ in this region). This is more pronounced in models exploded at lower energies. 
As expected, although heating from the radioactive decay chain of $\Ni$ extends the plateau and elevates 
the Nickel tail, it has very little impact on the final integrated value of $\ET$ calculated 
from our model light curves. Indeed, the variations of $\ET$ for the same explosion but different $\Ni$ are 
at a few per-cent level. 

\begin{figure}
\centering
\includegraphics[width=\columnwidth]{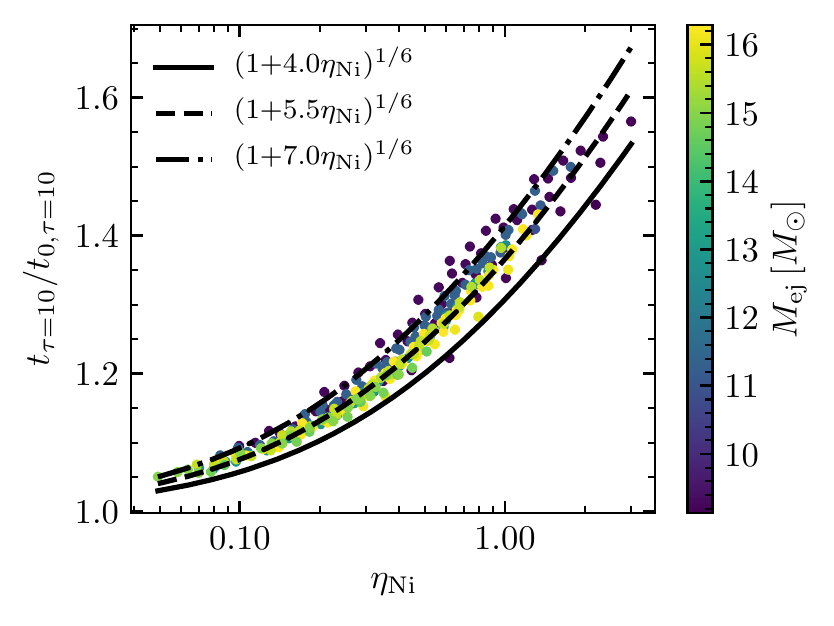} 
\caption{Plateau duration for models with $\Ni$ relative to the same explosion with no $\Ni$, 
versus the parameter $\etaNi$. Color corresponds to ejecta mass, and can be used to distinguish 
between the individual progenitor models. Black lines correspond to the plateau extension 
factor $(1+a\etaNi)^{1/6}$ for different values of $a$.}
\label{fig:etaNi}
\end{figure}

Dimensionally, using the Popov scalings for $\Lfifty$ and plateau duration with no $\Ni$ ($\tnone$), 
$\ET$ is expected to scale as 
\begin{equation}
\ET\propto \Lfifty \tnone^2 \propto \Mej^{1/2} \Eexp^{1/2} R,
\label{eq:NakarScaling}
\end{equation}
and thus $\etaNi$ should scale as $\MNi/\ET$. A more detailed derivation of this same 
scaling is given in \citet{Shussman2016}. This recovers the extension to the 
plateau duration given by Equation \eqref{eq:KasenScaling}, recast as 
\begin{equation}
\ttau/t_0 = (1 + a \etaNi)^{1/6},
\label{eq:KozyrevaScaling}
\end{equation}
where the scaling factor $a$ can be fit from models and 
encodes information about the internal structure of the ejecta, and in particular the 
concentration of $\Ni$. 
\citet{Kozyreva2018} find that for typical models, 
$a\approx4$ (their Figure 5). Figure \ref{fig:etaNi} shows the extension
of the plateau as a function of $\etaNi$ in our models.
We find slightly higher values for $a$, with more models falling along $a\approx5.5$, 
indicating a larger impact of $\Ni$ on the plateau 
duration, in part because $\etaNi$ encodes information about $\Ni$ mixing, and our models 
make use of the Duffell RTI prescription whereas mixing is parameterized in \citet{Kozyreva2018}. 
We show good agreement with the functional form in Equation \eqref{eq:KozyrevaScaling}. 

For SNe with a reasonably well-sampled bolometric light curve where $\MNi$ is measured
from the Nickel tail, $\ET$ can be calculated and used to constrain $\Mej$, $\Eexp$, 
and progenitor $R$ for a given explosion. In addition, 
$\ET$ can provide a critical constraint for explosions with lower $\MNi$, where the $\Ni$ decay chain does 
not dominate the internal energy of ejecta and thus the simple power law of Equation \eqref{eq:tptscaling} 
should not apply. Although observationally $\MNi$ must be extracted from the Nickel tail in order 
to calculate $\ET$, $\ET$ does not follow any scaling with $\MNi$, as it subtracts 
the contribution of $\Ni$ heating in the light curve evolution.

\begin{figure}
\centering
\includegraphics[width=\columnwidth]{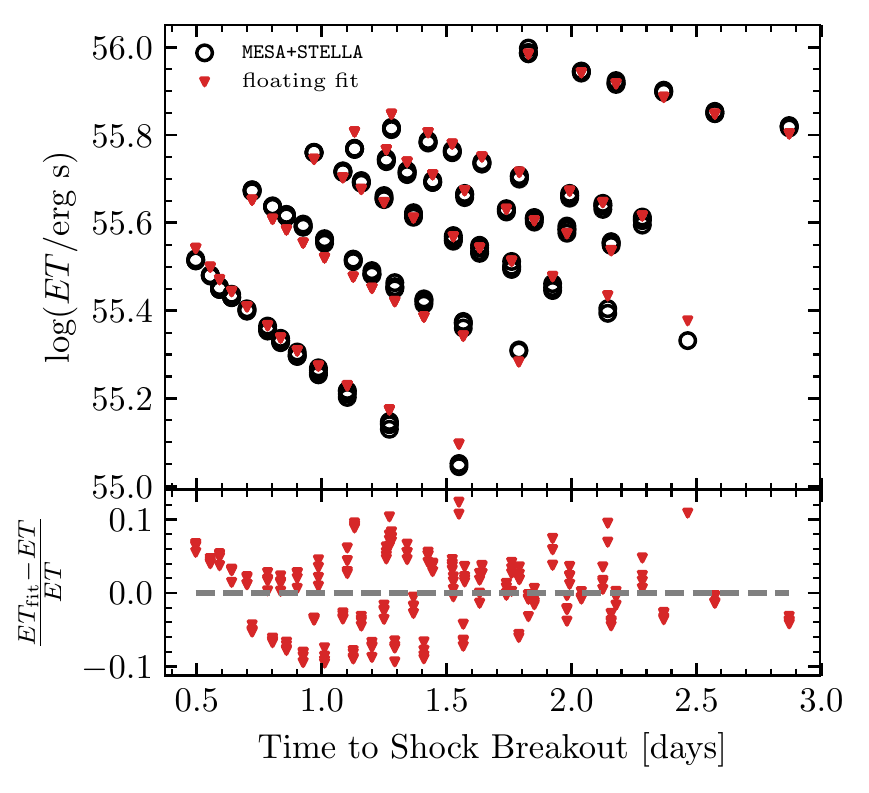} 
\caption{Calculated values of $\ET$ for 218 SN light curves and the values recovered with 
the fitted Equation \eqref{eq:ETscaling} (upper panel), and fractional deviations between $\ET$ in the 
models and recovered using Equation \eqref{eq:ETscaling} (lower panel).}
\label{fig:UglyPlotET}
\end{figure}

To determine how $\ET$ scales with $\Mej$, $\Eexp$, and $R$ in our models, 
we use the same 218 model light curves as with $\tpt$ in Equation \eqref{eq:tptscaling}, to recover
\begin{align}
\begin{split}
\log(\ET/{\rm erg\,s}) = & 55.460 + 0.299\log(\Mten) \\&+  0.435 \log(\Efoe) + 0.911\log(\Rfh)
\label{eq:ETscaling}
\end{split}
\end{align}
for our suite of models, which does not include interactions CSM. 
This scaling has a slightly shallower dependence on $\Mej$, $\Eexp$, and progenitor $R$ 
than Equation \eqref{eq:NakarScaling}.
The agreement between our models and Equation \eqref{eq:ETscaling} is shown in Figure \ref{fig:UglyPlotET}. 
 RMS deviations between our models are 5.0\%, with a maximum deviation of 12.4\%. 
Although the fit was performed on models with $\MNi\gtrsim0.03\Msun$ to be consistent with our 
set of models for $\tpt$, the recovered scaling applies similarly well for our models with $\MNi<0.03\Msun$, 
with  RMS deviations of 5.3\% and all deviations under 20\%. 
The overlapping black rings in Figure \ref{fig:UglyPlotET} show the typical scatter in values of $\ET$ 
for the same explosion when varying $\MNi$. Each set of overlapping rings corresponds to $\ET$ for 
a single progenitor model exploded with a single $\Eexp$, but with different values of $\MNi$. 
This scatter in $\ET$ when only varying $\MNi$ is well within the scatter between the models 
and the fitted Equation \eqref{eq:ETscaling}. 

\section{Observed Velocity Evolution\label{sec:VELOCITIES}}
We now discuss the diagnostic value of the material velocity inferred from the absorption minimum 
of the \FeA\ line, often measured and reported at day 50, $\vFef$. Ideally, the measured Fe line
velocities would provide an additional quantitative measurement that would allow for estimation of 
progenitor and explosion properties \citep{Pejcha2015a, Muller2017}. However, as we show here, these 
measurements are highly correlated with bolometric luminosity measurements at a fixed time on the
plateau, and are largely redundant at day 50. If there is no substantial CSM around the star, 
than earlier time ($\lesssim 20$ day) measurements may prove more useful (see Section \ref{sec:EARLYV}). 

The \FeA\ velocity is typically used to approximate the velocity at the photosphere ($\vPh$), 
although there is substantial evidence that measured line velocities are typically higher than that
predicted for the model photosphere ($\tau=2/3$) (e.g. \citealt{Utrobin2017}; 
MESA IV). In a homologously expanding medium, the strength of a given 
line is quantified using the Sobolev optical depth \citep{Sobolev1960,Castor1970,Mihalas1978,Kasen2006}, 
which accounts for the shift in the line profile due to the steep velocity gradient in the
ejecta. This is captured in \MESA+\STELLA\ following \mesafour, 
where the $\tauS=1$ condition is used to measure iron line velocities 
($\vFe$). Although in the following we discuss both this velocity and the velocity 
at the model photosphere, we recommend using $\vFe$ defined when $\tauS=1$ 
when comparing to observations. 

\subsection{Velocities in Explosion Models}
When the velocity profile of the ejecta becomes fixed in time,
this material is said to be in homologous expansion. Analytically, homology 
is often approximated $r=vt$ for a fluid element at radial coordinate $r$ with 
velocity $v$ at time $t$. While not quite true for material in the center of the ejecta, which 
is expanding more slowly and therefore the initial radial coordinate is still relevant, 
this approximation generally holds for faster-moving material which has experienced 
more significant expansion at a given time, as well as for the slower-moving material 
at late times when it is becoming visible. This is reflected in Figure 32 of MESA IV. 

\begin{figure}
\centering
\includegraphics[width=\columnwidth]{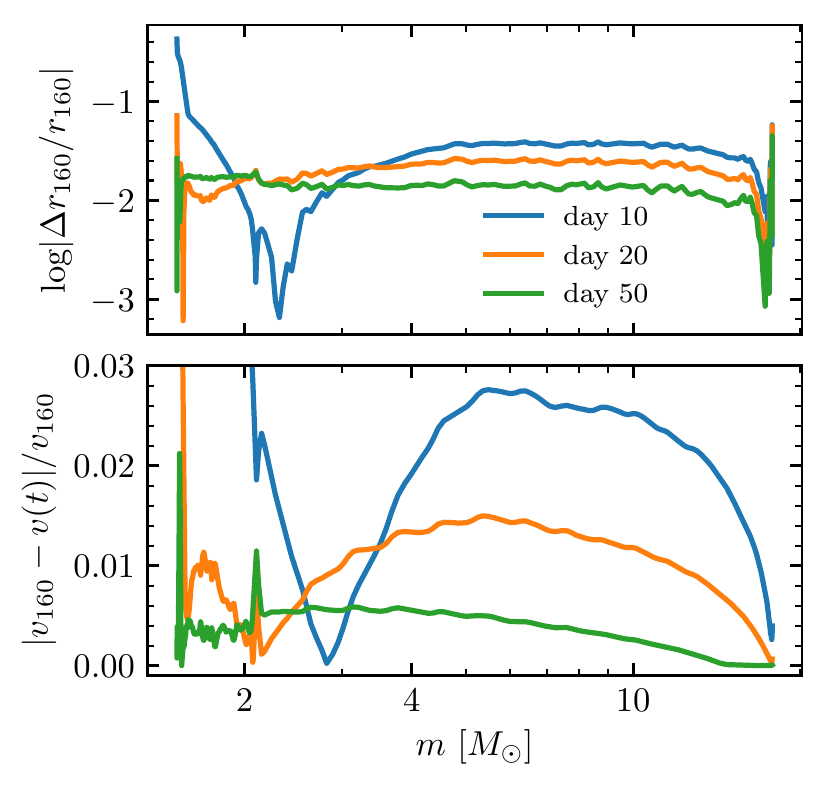} 
\caption{Agreement between hydrodynamical models and homologous expansion.
The upper panel shows the relative error in predicting the radial profile day 160 
by assuming homology starting at days 10, 20, and 50, compared to the true radial coordinate at 
day 160 in \STELLA. The lower panel shows the absolute relative deviation between the
velocity profiles at days 10, 20, and 50, and the profile at day 160.}
\label{fig:homology}
\end{figure}

Many software instruments devoted to modeling radiative transfer, such as \Sedona, 
assume homologous expansion in the true sense of a fixed velocity profile. 
Figure \ref{fig:homology} shows the extent to which this is satisfied in our M16.3\_R608 model exploded with $\foe$. 
The upper panel shows the relative error in predicting the radial coordinate of a fluid element at day 160 
by assuming $r_{\rm h}({\rm day\ 160})~=~r(t_0)~+~v(t_0)(160\days-t_0)$ for homology starting at 
$t_0$ = days 10, 20, and 50. We define $\Delta r_{160}=r_{160}-r_{\rm h}(\rm day 160)$, 
where $r_{160}=$ the true radius of that fluid element at day 160 in \STELLA.
The lower panel shows the deviation between the velocity profiles at days 10, 20, and 50, and that at day 160. 
Before homology, the innermost material is moving slightly faster than its day 160 value, 
and the outer material is moving slightly slower. Even in the envelope, there is deviation 
between the day 10 velocity profile and day 160 at the level of a few per-cent. 
By day 20 this falls below 2\%, and by day 50 typical deviations of the velocity profile 
in the bulk of the ejecta from the velocity profile at day 160 are at the level of 0.5\%. 
Generally, by day 20, the difference in predicted radial 
coordinate of the half-mass fluid element at day 160 is below 3\% of its true value in the 
hydrodynamical simulation. At this time the radial coordinate predicted for day 50 is also 
within 2\% of its true value at day 50. 

\begin{figure}
\centering
\includegraphics[width=\columnwidth]{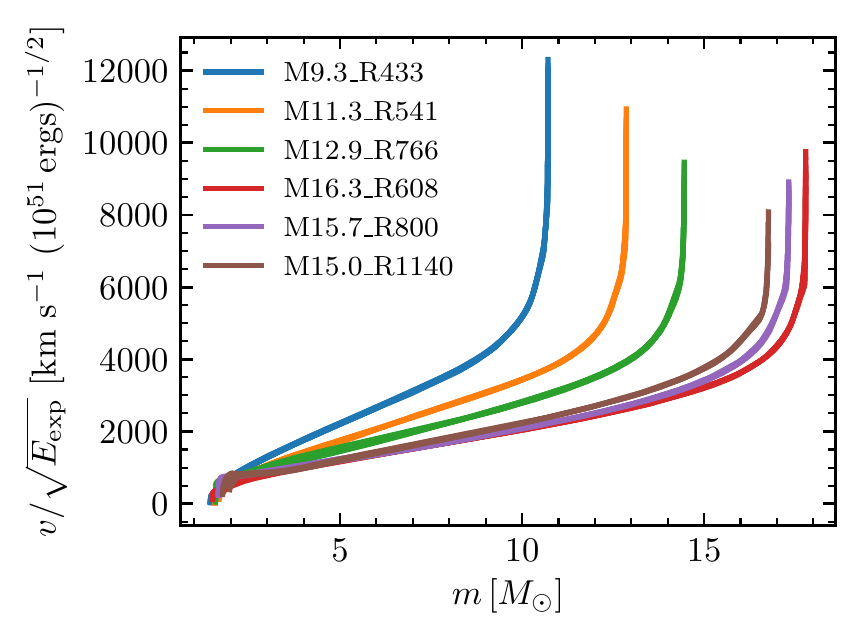} 
\caption{Day 50 velocity profiles in \STELLA, divided by 
the square root of $\Eexp$, versus mass coordinate for 6 unique progenitor
models with a variety of explosion energies. Each colored line represents a different 
model, and lies on top of a collection of $6-12$ nearly identical lines which correspond to 
different explosion energies for the same progenitor model.}
\label{fig:vday50masses}
\end{figure}

\begin{figure}
\centering
\includegraphics[width=\columnwidth]{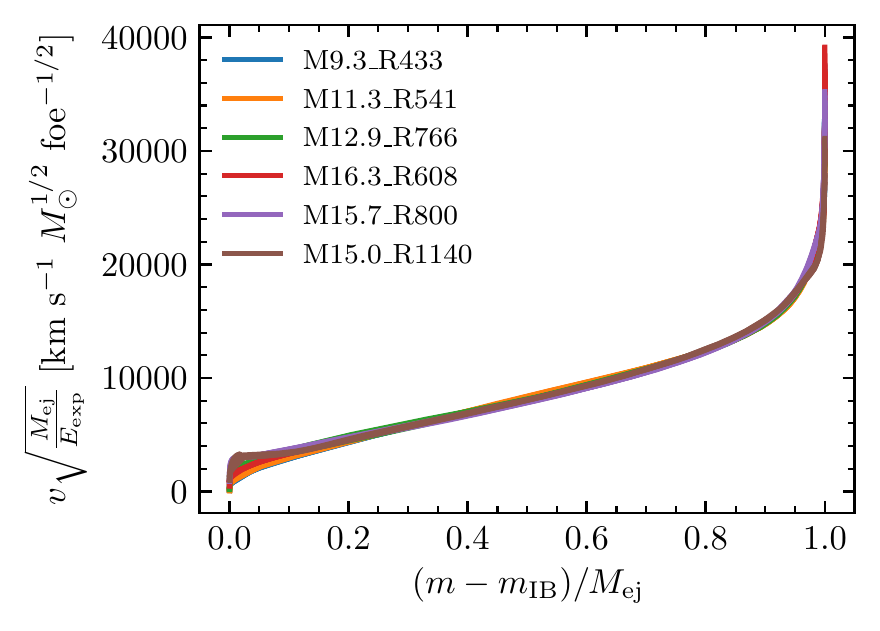} 
\caption{Day 50 velocity profiles in \STELLA, multiplied by 
$\sqrt{\Mej/\Eexp}$, versus dimensionless ejecta mass coordinate for all 57 explosions
with energies sufficient to yield little fallback. For a fixed fractional
position within the ejecta, velocities obey the scaling in Equation \eqref{eq:badvscaling}.}
\label{fig:vRootMdivE}
\end{figure}

Figure \ref{fig:vday50masses} shows approximately homologous velocity profiles (taken here at day 50) 
scaled by the square root of $\Eexp$ for all 6 progenitor models at 
all energies that cause sufficiently little fallback. 
Each family of colored lines reflects explosions of an individual model, and each of the 6 families of lines contains
the profiles for multiple explosion energies for that model. When looking at any
fixed mass coordinate within a single progenitor model, the fluid velocity divided by $\sqrt{\Eexp}$ is constant.  
Moreover, as shown in Figure \ref{fig:vRootMdivE}, looking at the same dimensionless mass 
coordinate inside the ejecta and scaling also by the square root of $\Mej$, 
this relationship holds for any dimensionless ejecta mass coordinate throughout the entire 
velocity profile, with small variations only near the inner boundary, where 
the reverse shock becomes relevant and where fallback has a greater effect. 

\begin{figure}
\centering
\includegraphics[width=\columnwidth]{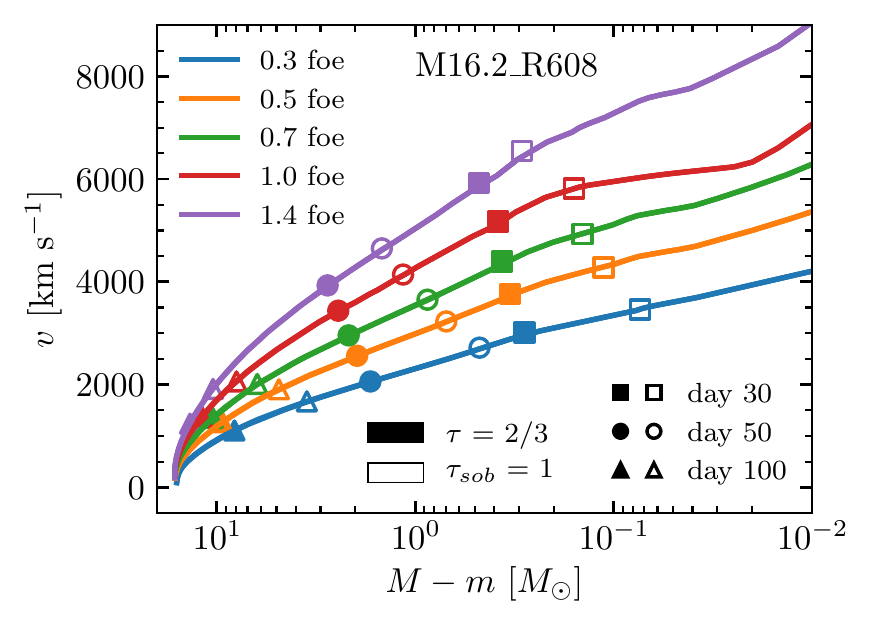} 
\caption{Day 50 velocity profiles versus overhead mass coordinate in \STELLA\ for our 
M16.3\_R608 model at 5 different explosion energies. We have included photospheric 
(filled markers) and  \Feline\ velocities (open markers) for day 30, 50, 
and 100 as denoted by shape of the marker.}
\label{fig:vday50constantmass}
\end{figure}

\citet{Popov1993}, \citet{Pejcha2015a}, and others, have often assumed that
\begin{equation}
\Eexp \approx \frac{1}{2} \Mej \vPhf^2,
\label{eq:badvscaling} 
\end{equation}
where $\vPhf$ is the photospheric velocity at day 50,
in order to close the system of equations for $\Lfifty$ and $\tpt$ as a function of 
$\Mej$, $\Eexp$ and progenitor radius $R$. 
While the scaling law suggested in Equation \eqref{eq:badvscaling} holds for the fluid 
velocity at a fixed dimensionless ejecta mass coordinate, as shown in Figure \ref{fig:vRootMdivE}, 
as the photosphere moves deeper into the ejecta, it does not probe velocities at the same mass 
coordinate at a given time post shock-breakout. Rather, at a fixed time in the
evolution, faster-expanding ejecta in higher energy explosions allows the observer to see deeper mass 
coordinates, compared to a lower energy explosion of the same star. This is evident in 
Figure \ref{fig:vday50constantmass}, which shows velocity profiles for the M16.3\_R608 model at 5 
different explosion energies, marking the location of the photosphere and \Feline\ at fixed times.
As a result of the expanding ejecta, we expect a shallower scaling for velocity 
as a function of energy at fixed mass than the naive $\vPhf \propto \Eexp^{1/2}$.
Indeed a linear fit for a single model with fixed ejecta mass and radius finds shallower scalings:
$\vPhf \propto \Eexp^{0.36}$, and $\vFef \propto \Eexp^{0.30}$. 
These scalings approximately hold for the other individual models.

\begin{figure}
\centering
\includegraphics[width=\columnwidth]{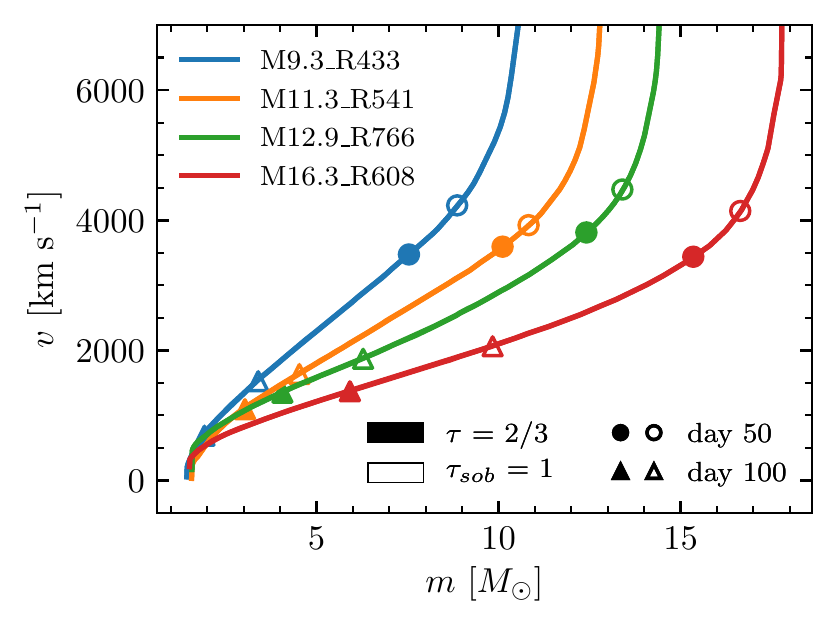}  
\caption{Day 50 velocity profiles for 4 different models all exploded with an 
energy of $1.0\times 10^{51}$ ergs. Filled markers indicate the location of the photosphere and 
open markers indicate where the \Feline\ is formed. Velocities at days 50 and 100 are denoted by 
the shape of the marker.}
\label{fig:vday50constantE}
\end{figure}

Additionally, a simple velocity scaling with $\Mej$ and $\Eexp$ becomes murkier when comparing 
across models of different masses at fixed explosion energy, since there is no 
reason for the same explosion energy to yield the 
``same" mass coordinate at the same time in two different progenitors. 
In fact, as seen in Figure \ref{fig:vday50constantE}, $\vFe$ and $\vPh$ at day 50 are not 
even monotonic in $\Mej$ for different stars at fixed $\Eexp$. Thus, we cannot 
derive any power law for $\vPhf$ or $\vFef$ solely as a function of $\Mej$ and $\Eexp$. 
As we show in the following section, additional dependences are relevant 
(Equation \ref{eq:vPhfloating}). 

\subsection{$\Lfifty$ - $\vPhf$ Relation}
This result highlights a true degeneracy, discovered observationally by 
\cite{Hamuy2003} and explained by \citet{Kasen2009}.
We start with the Stefan-Boltzmann formula for luminosity, 
$L = 4 \pi R_{\rm phot}^2 \sigma \Teff^4$, where $R_{\rm phot}$ is the photospheric radius, 
and note that $\Teff$ is roughly constant 
at the photosphere and set by H recombination to $\Teff \approx$ 6000\,K. 
At fixed time on the plateau, (e.g. day 50) while the ejecta is expanding homologously with radial 
position $r \approx v t$ for any given mass coordinate, for the photosphere at day 50 
$L_{50} \propto \vPhf^2$ and so $\vPhf \propto \sqrt{\Lfifty}$.
In this way, the luminosity, together with homologous expansion, sets the location of the photosphere 
within the expanding ejecta, which in turn sets the velocity measured at or near the photosphere.

\begin{figure}
\centering
\includegraphics[width=\columnwidth]{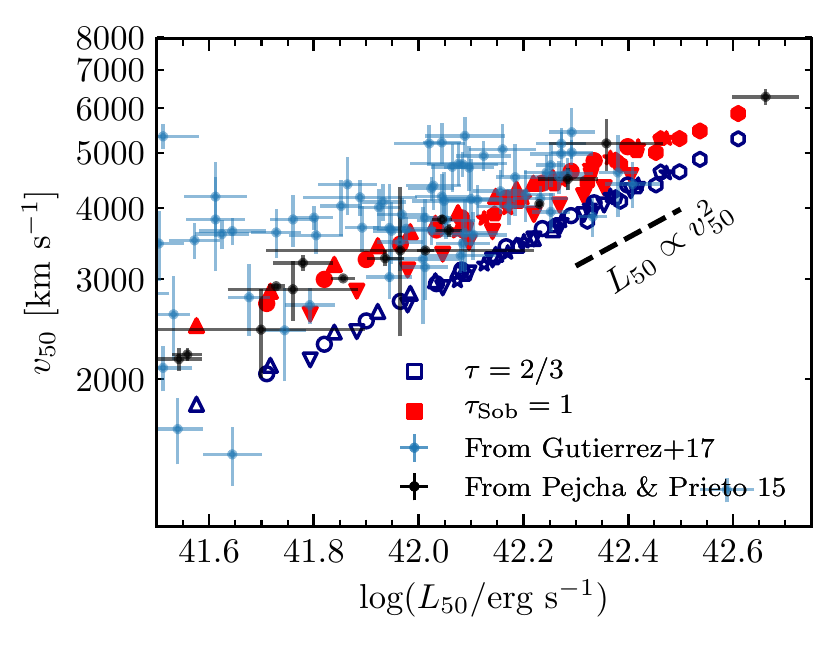} 
\caption{Velocity versus Luminosity at day 50 for a variety of progenitor models 
and explosion energies. Open navy blue markers denote photospheric velocities ($\tau=2/3$) 
and closed red markers denote Fe II 5169 \AA\ velocities ($\tauS=1$). Points with error bars
are data from 2 samples: \citet{Gutierrez2017} (blue) and \citet{Pejcha2015b} (black).}
\label{fig:v50vsL50}
\end{figure}

Figure \ref{fig:v50vsL50} shows $\vPhf$ and $\vFef$ versus $L_{50}$ for all 57 explosions 
which experience sufficiently little fallback (6 models with 6-12 explosion energies each).
Also plotted are data from \citet{Pejcha2015b} and \citet{Gutierrez2017}.\footnote{Luminosities 
from \citet{Pejcha2015b} are bolometric luminosities provided by O. Pejcha (private 
communication). Luminosities from \citet{Gutierrez2017} were estimated from $M_V$ measurements at 
day 50 provided by C. Gutierrez (private communication), assuming negligible bolometric correction 
BC$\approx0$ for $M_V$ following the correction for SN1999em on the plateau, shown in 
\citet{Bersten2009}. Typical V band bolometric corrections on the plateau of Type IIP SNe are 
BC$\approx -0.15$ to $0.1$, and  the variation in $\log{L_{50}}$ from assuming a BC of $0$ versus 
other values within that range is smaller than the error bars on the data.}
In both observational data sets, velocities are inferred from the \Feline, 
suggesting that these velocities are better captured in our models 
at $\tauS\approx1$, rather than assuming the line is formed at the 
photosphere ($\tau=2/3$). We also see good agreement between our models 
and the scaling $\vPhf \propto \sqrt{\Lfifty}$. 

\begin{figure}
\centering
\includegraphics[width=\columnwidth]{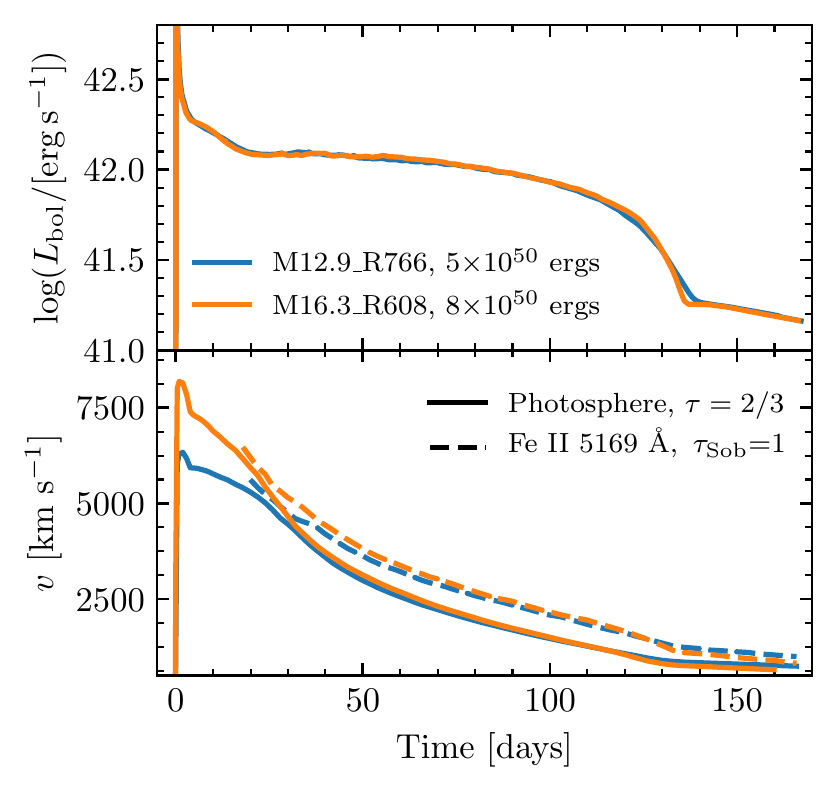} 
\caption{Luminosity and velocity for the 99em\_16 (renamed  M12.9\_R766) and 99em\_19 
(renamed M16.3\_R608) models from \citet{Paxton2018} with different explosion energies in order to 
attain light curve agreement.}
\label{fig:99emMatching}
\end{figure}

It is therefore unsurprising that the Fe velocities during the plateau 
phase match the data for a model with a luminosity match at day 50. 
This was seen in Section 6 of MESA IV, Figure 42, where two 
models with light curve agreement with SN199em show identical velocity evolution. 
Figure \ref{fig:99emMatching} shows the luminosity and velocity of those two progenitor models, 
renamed M12.9\_R766 and M16.3\_R608 in our suite, blown up with slightly adjusted explosion energies 
to produce even better light curve agreement. In the case where models match closely in both $L$ and 
$t_{p}$, the agreement in velocity is excellent throughout the evolution of the SN. 

As with $L_{50}$ in Section \ref{sec:LUMINOSITIES}, we fit a power law for $v_{50}$ as a function
of $\Mej$, $\Eexp$, and $R$ to our models with constant nickel mass $\MNi=0.03\Msun$. We do this 
with both $\vFef$ and $\vPhf$ at day 50, noting that observationally, $\vFef$ is the relevant scaling. 
For the photospheric velocity at day 50, we found power laws that are very similar
to the scaling found if $\vPhf\propto \Lfifty^{1/2}$:
\begin{align} 
\begin{split}
&\log(\vPh/{\rm km\ s^{-1}}) = \\
&3.54 - 0.19 \log M_{10} + 0.36 \log E_{51} + 0.32 \log R_{500},
\end{split}
\label{eq:vPhfloating}
\end{align}
where the prefactor and power law coefficients are all fit from our models.

\begin{figure}
\centering
\includegraphics[width=\columnwidth]{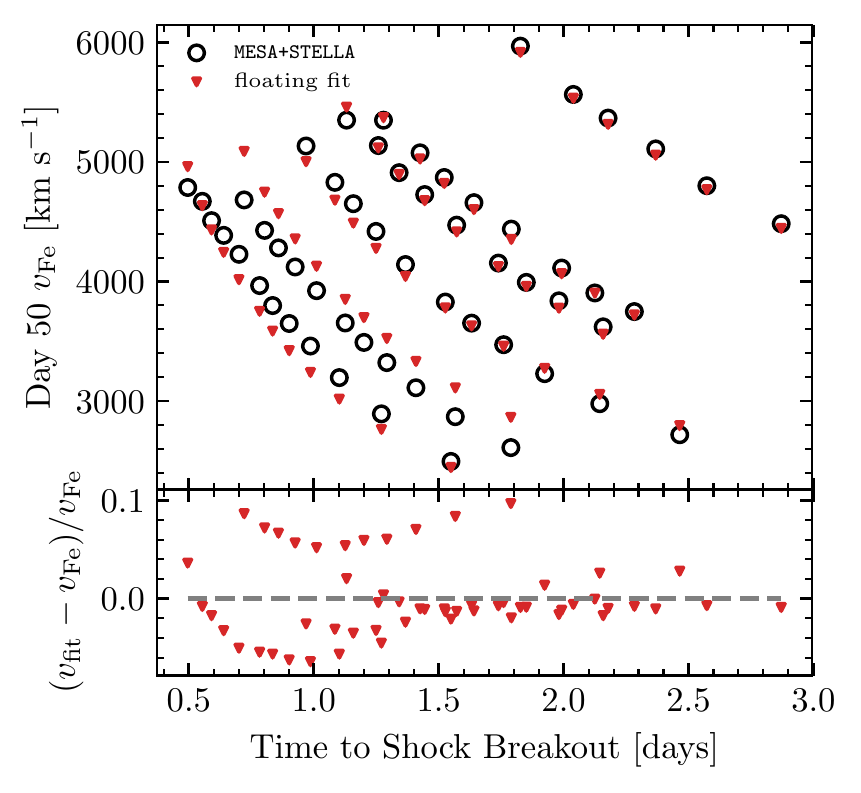} 
\caption{Model $v_{{\rm Fe}}$ ($\tauS=1$) at day 50, and velocities recovered 
with the fitting formulas (upper panel), and their residuals (lower panel) for our suite of 57 explosions 
with $\MNi=0.03\Msun$. So as to clarify the model space, we use time to shock breakout as the x-axis.}
\label{fig:v50fits}
\end{figure}

This is valuable insofar as it reinforces the degeneracy 
highlighted in Figures \ref{fig:v50vsL50} and \ref{fig:99emMatching}, but, as discussed, this velocity is unmeasurable, and 
observed \Feline\ velocities are better estimated by ($\tauS=1$). 
A similar fit to $\vFe$ at day 50,
\begin{align} 
\begin{split}
&\log(\vFef/{\rm km\ s^{-1}}) = \\
&3.61 - 0.12 \log M_{10} + 0.30 \log E_{51} + 0.25 \log R_{500},
\end{split}
\label{eq:vfloating}
\end{align}
yields higher predicted velocities everywhere, and shows somewhat shallower dependence on each of the 
explosion properties.
The model Fe line velocities and their residuals as compared with
Equation \eqref{eq:vfloating} are shown in Figure \ref{fig:v50fits}. 

Although the degeneracy is less pronounced for $\tauS=1$ than for the photosphere, 
with some scatter in Figure \ref{fig:v50vsL50} and differences in the recovered power laws, 
this scatter is small compared to intrinsic variations in luminosity and plateau 
duration, and is therefore insufficient to break the degeneracy between $v$ and $L$
in order to provide accurate estimates for $\Mej$, $\Eexp$, and $R$. 
It is for this reason that we do not advocate using measured velocities at day 50 to infer 
explosion properties.

\section{Families of Explosions\label{sec:PREDICTIONS}}
\subsection{Inverting Our Scalings}
Due to the degeneracies highlighted in Section \ref{sec:VELOCITIES}, 
we cannot simply extract $\Mej$, $\Eexp$, and $R$ from light curve measurements and $\vFef$. 
Attempting to invert all three scalings (Equations \eqref{eq:LFloatingFit}, \eqref{eq:ETscaling}, 
and \eqref{eq:tptscaling}) is ill-conditioned and within the scatter within our models. However,
we can use the scalings to solve for two of the three relevant 
explosion properties as a function of the third, revealing a family of possible explosions 
that yield nearly identical bolometric light curves. 

SNe with direct progenitor observations are improving with time, so we solve 
Equations \eqref{eq:LFloatingFit} and \eqref{eq:tptscaling}
for $\Mej$ and $\Eexp$ as a function of $\MNi$, $\Lfifty$, $\tpt$, and $R$, to find
\begin{equation}
\begin{split}
\begin{aligned}
\log(\Efoe) &= -0.728 + 2.148 \log(\Lft) - 0.280\log(\MNi) \\&+ 2.091\log(\tptwo) - 
    1.632\log(\Rfh), \\
\log(\Mten) &= -0.947 + 1.474 \log(\Lft) - 0.518 \log(\MNi) \\&+ 3.867 \log(\tptwo) - 
    1.120 \log(\Rfh),
\label{eq:EandMofR}
\end{aligned}
\end{split}
\end{equation}
where $\MNi$ is in units of $\Msun$, $\Lft = \Lfifty/10^{42}$ erg s$^{-1}$ 
and $\tptwo=\tpt/100\,\days$. 
Alternatively, we can use a measured $\ET$ rather than $\tpt$ to find 
\begin{equation}
\begin{split}
\begin{aligned}
\log(\Efoe) &= -0.587 - 1.497\log(\Rfh) \\&+ 1.012\log(\ETff) + 0.756 \log(\Lft), \\
\log(\Mten) &= -0.685 - 0.869 \log(\Rfh)  \\&+ 1.872 \log(\ETff) - 1.101 \log(\Lft),
\label{eq:EandMofR_ET}
\end{aligned}
\end{split}
\end{equation} 
where $\ETff=\ET/10^{55} {\rm erg\, s}$.

\begin{figure}
\centering
\includegraphics[width=\columnwidth]{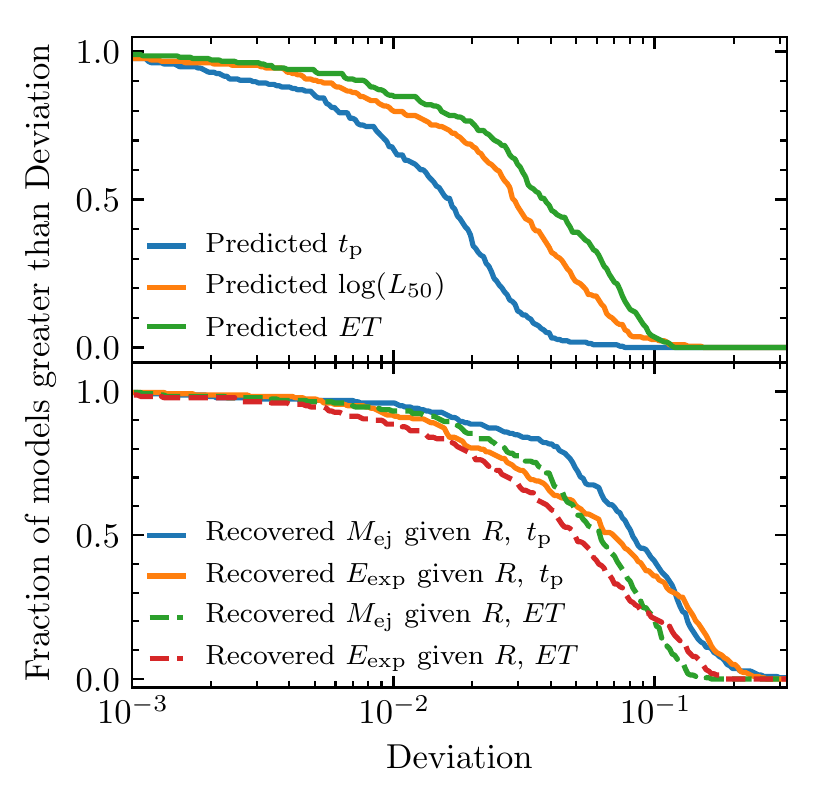} 
\caption{Distribution of deviations between our models and Equations \eqref{eq:LFloatingFit} and 
\eqref{eq:tptscaling} (upper panel), and the distribution of deviations between recovered 
values of $\Mej$ and $\Eexp$ by applying the inverted scalings (Equation \eqref{eq:EandMofR}) 
to the model light curve properties and radii, and the models themselves (lower panel). 
This gives a heuristic for the agreement between the fitted 
formulae and our suite of models.} 
\label{fig:DeviationFractions}
\end{figure}

Before demonstrating how to apply these fitting formula to observed SNe, we show how well 
modeled events can be matched. The upper panel of Figure \ref{fig:DeviationFractions} shows the fraction of 
models with light curve properties matching their fitted values (applying Equations \eqref{eq:LFloatingFit},
\eqref{eq:tptscaling}, and \eqref{eq:ETscaling}) within a given deviation tolerance shown on the x-axis. 
The lower panel shows the fraction of models in which we can recover 
the values of $\Mej$ and $\Eexp$ within a given deviation tolerance 
by applying Equation \eqref{eq:EandMofR} (solid lines) or Equation \eqref{eq:EandMofR_ET} 
(dashed lines) to the model light curve observables and $R$. 
Given that there is no statistical meaning to the sample of models beyond
probing different regions of parameter space, this merely provides a heuristic 
guide to how well our sample of models match with the fitted formulae.

Applying Equation \eqref{eq:EandMofR} using $\tpt$ to our suite of Nickel-rich SNe, 
we recover $\Mej$ and $\Eexp$ with  RMS deviations between the models and the fits of 10.7\% and 10.4\%, 
respectively, with maximum deviations of 35\% and 27\%. 
Using $\ET$ and Equation \eqref{eq:EandMofR_ET}, we recover $\Mej$ and $\Eexp$ 
with  RMS deviations between the models and the fits of 7.3\% and 7.6\%, 
respectively, with maximum deviations of 16\% and 18\%. 
Although the modeling uncertainties for the inverted $\ET$ scalings are smaller than 
those which use $\tpt$, the observable uncertainty is greater and may be accompanied by an offset, 
as excess emission within the first 10-40 days due to interaction with CSM may cause
an excess in $\ET$ as compared to our models.

\begin{figure}
\centering
\includegraphics[width=\columnwidth]{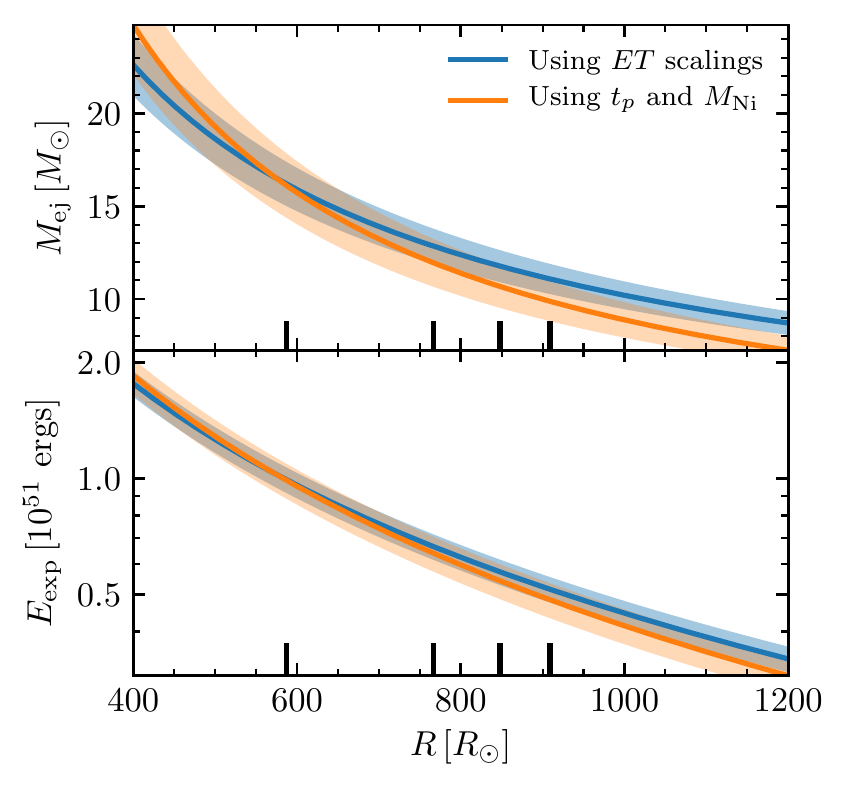} 
\caption{Degeneracy curves applying Equation \eqref{eq:EandMofR} (orange) and
Equation \eqref{eq:EandMofR_ET} (blue) as a function of progenitor radius $R$
for an ``observed" SN with $\log(\Lfifty/{\rm erg\,s^{-1}})$=42.13, 
$\log(\ET/{\rm erg\,s})$=55.58, $\tpt$=123, and $\MNi=0.045\Msun$.
The shaded region corresponds to the  RMS deviations between our models and 
the values recovered by applying Equations \eqref{eq:EandMofR} and \eqref{eq:EandMofR_ET}. Short black lines
correspond to the radii of the M17.8\_R587, M12.9\_R766, M10.2\_R848, and M9.8\_R909 models, which produce the 
light curves, velocities, and $\ET$ evolution shown in Figure \ref{fig:DegenerateLCs}.} 
\label{fig:WedgeDiagram1}
\end{figure}

Using these relations, we now show how very comparable light curves
(and thus comparable \Feline\ velocities on the plateau) can be produced with different 
progenitors exploded at different energies. 
Figure \ref{fig:WedgeDiagram1} shows an example of the family of models in $\Mej-\Eexp$ parameter 
space as a function of $R$ that could produce an ``observed" SN light curve 
with $\log(\Lfifty/{\rm erg\,s^{-1}})$=42.13, $\log(\ET/{\rm erg\,s})$=55.58, $\tpt$=123, and $\MNi=0.045\Msun$,
which are the values matching a randomly selected model out of our suite: 
the M12.9\_R766 model exploded with $\Eexp=6\times10^{50}$ ergs and that $\MNi$. 

\begin{figure}
\centering
\includegraphics[width=\columnwidth]{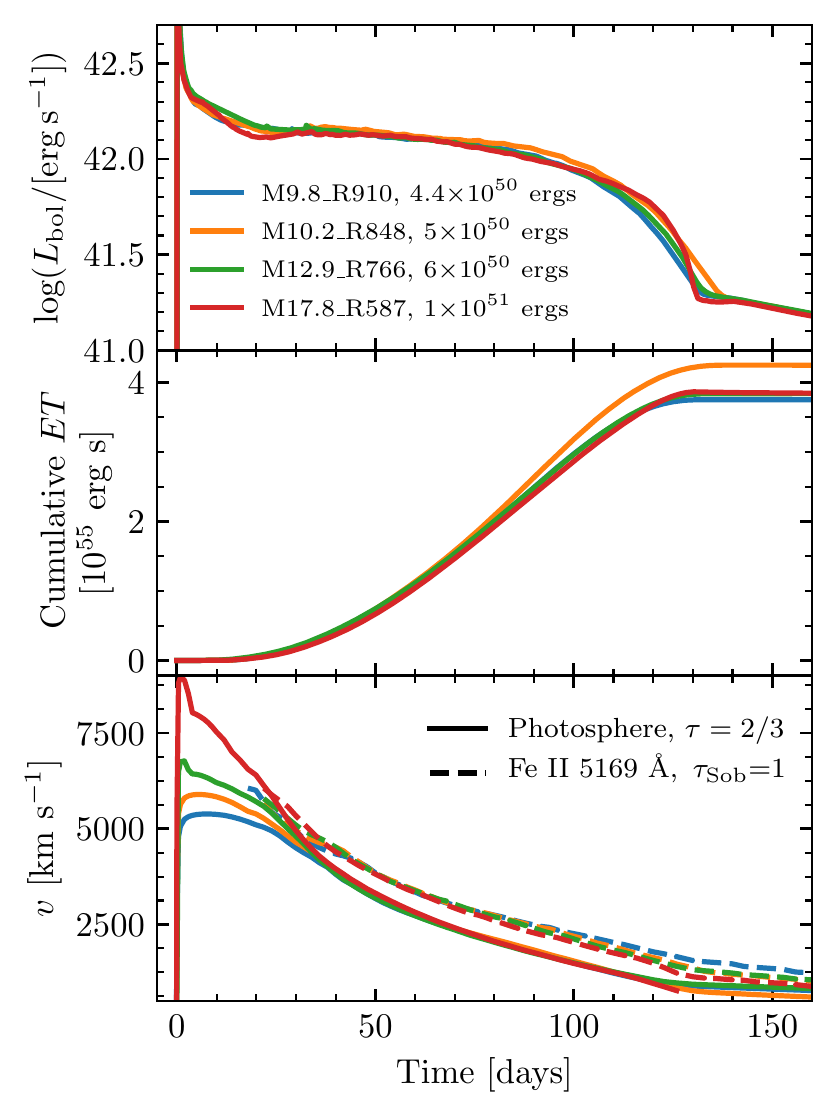}
\caption{Light curves, cumulative ET, and velocities for four different explosions which 
yield nearly the same $\Lfifty$, $\tpt$, and velocities at day 50.} 

\label{fig:DegenerateLCs}
\end{figure}

To exhibit how this exercise would proceed, we constructed three additional models consistent with the bands in Figure
\ref{fig:WedgeDiagram1}, based off Equation \eqref{eq:EandMofR} using $\tpt$. 
We then explode these progenitor models with $\Eexp$ as dictated by 
the degeneracy curve. 
We created multiple such models: one with $\Mej=17.8\Msun$ and 
$R=587\Rsun$, which we explode with $1\times10^{51}$ ergs, one with $\Mej=10.2\Msun$ and 
$R=848\Rsun$, which we explode with $5\times10^{50}$ ergs, and one with $\Mej=9.8\Msun$ and 
$R=909\Rsun$, which we explode with $4.5\times10^{50}$ ergs. The values of $R$ for these three models, and for M12.9\_R766 exploded with $6\times10^{50}$ ergs,
are shown as black tick marks in Figure \ref{fig:WedgeDiagram1}.
Figure \ref{fig:DegenerateLCs} shows the resulting light curves, velocities, and accumulated ETs. 
We see very good agreement in $\Lfifty$ and along the plateau, and recover $\tpt$ values from 
120 to 125 days for all four light curves. 

The values of $\ET$ for three of the four light curves agree within $\approx2$\%, 
ranging from 3.75 to 3.84$\times10^{55}$ erg s; however, the $5\times10^{50}$ erg explosion of the 
M10.2\_R848 model has a value of $\ET$ which is noticeably higher, at 4.26$\times10^{55}$ erg s.
Additionally, velocities agree on the plateau, and thus cannot be used to 
break the light curve degeneracy, which at least spans a factor of 2 in explosion energy, 
nearly a factor of 2 in $\Mej$, and a factor of 1.5 in progenitor $R$. This captures much of 
the parameter space in which IIP SNe from RSG progenitors could be produced to begin with! 

\subsection{The Importance of Velocities at Early Times \label{sec:EARLYV}}
Although velocity measurements at day 50 are largely degenerate with measurements of $\Lfifty$, as discussed
in detail in Section \ref{sec:VELOCITIES}, early time velocities up to day $\approx$20 could be used to 
distinguish between low-energy explosions of large-radius lower-mass RSGs and high-energy explosions 
of compact-radius high-mass RSGs in cases where there is minimal CSM present. 
As seen in the lower panel of Figure \ref{fig:DegenerateLCs}, higher energy explosions of compact stars 
yield faster velocities at early times. Before around day 20, the radial coordinate of the photosphere 
is moving outward, and the declining photospheric temperature is set by shock cooling rather than by 
recombination. Thus in this phase the velocity measured near the photosphere is not dictated 
by the plateau luminosity as it is at day 50. 
Early light curves and photospheric velocities are discussed in detail by \citet{Morozova2016} and \citet{Shussman2016b}.
\citet{Shussman2016b} find an expression for the photospheric velocity at early times as a function of $\Mej$, $\Eexp$, 
and $R$ (their Equation 48), assuming that the density profile of the progenitor model behaves like a power 
law in radial coordinates. After the photosphere leaves the so-called breakout shell ($5 \days \lesssim t \lesssim 20\days$),
\citet{Shussman2016b} find that $\vPh(t)/{\rm km\, s^{-1}} \approx 1.2\times10^4 
\Mfifteen^{-0.3} \Efoe^{0.38} \Rfh^{-0.14} \tday^{-0.2}$ where $\Mfifteen=\Mej/15\Msun$ and $\tday=t/\days$. 
At day 15 this equation describes our full suite of models with  RMS deviations of 5.5\% and with all 
deviations under 15\%.

As is also seen in Figure 2 of \citet{Morozova2016}, no single power law fully describes 
progenitor density profiles around the photospheric mass coordinate in our models 
for any fixed time in the light curve evolution.
Nonetheless our entire suite of models, which does not include the presence of circumstellar material, 
can approximately be described by the fitted power law
\begin{align}
\begin{split}
\log(\vPht) = &3.90 - 0.22\log(\Mten) \\& + 0.43 \log(\Efoe) - 0.13\log(\Rfh), 
\label{eq:v15scaling}
\end{split}
\end{align} 
where $\vPht$ is the photospheric velocity at day 15 in km~s$^{-1}$, with  RMS deviations of 3.7\% and a maximum
deviation of 10\% between the models and Equation \eqref{eq:v15scaling}. The dynamic range in $\vPht$ 
in our models is a factor of $\approx 3$, ranging from $\approx4,000-12,000$ km s$^{-1}$.

Although Equation \eqref{eq:v15scaling} and Shussman's Equation 48 describe our 
models well, we warn the reader that velocities at this time are sensitive to the 
density structure of the outermost ejecta including any asphericity, 
as well as any interactions with any circumstellar material present. 
Thus more work is needed in order to faithfully capture 
the early-time velocities and their dependence on the relevant properties 
of the explosion, especially in cases where CSM is present. 
Nonetheless, early time velocity measurements could in principle provide a third constraint and 
break the light curve degeneracies, thus allowing an inference of
$\Mej$, progenitor $R$, and $\Eexp$ for a given observed Type IIP SN.

\section{Concluding Remarks\label{sec:Conclusions}}

We have shown the utility of using \MESA+\STELLA\ to model an ensemble of Type IIP SN progenitors, 
a capability introduced by \citet{Paxton2018}. We introduced new best-fit scaling 
laws for the plateau luminosity at day 50, $\Lfifty$ (Equation \ref{eq:LFloatingFit}), 
and for the duration of the plateau $\tpt$ in the limit of Nickel-rich ($\MNi\geq0.03\Msun$) 
events (Equation \ref{eq:tptscaling}) as a function of ejecta mass, explosion energy, and progenitor radius. We also recovered a 
similar fit for the observable $\ET$ (Equation \ref{eq:ETscaling}). Velocity measurements on the plateau cannot 
be simply described by $\vPhf\approx(2\Eexp/\Mej)^{1/2}$ assumed by \citet{Popov1993} 
or the scaling given in \citet{Litvinova1983}, 
but rather scale with $\Lfifty$ as noted by \citet{Hamuy2003,Kasen2009} and others, 
shown in our Figure \ref{fig:v50vsL50}.  
While early-time velocities observed during the photospheric phase 
($\approx$~day~15) could provide a promising third independent constraint on
$\Mej$, $\Eexp$, and $R$, these velocities can be
affected by interaction with CSM, deviations from spherical symmetry, and
the specifics of the density profile of the progenitor star. 
Thus early velocities require more work in order to simply interpret in 
observed systems. Presently, given a bolometric light curve, 
one can at best recover a family of explosions which produce comparable 
light curves and thereby velocities on the plateau, as demonstrated in 
Figures \ref{fig:99emMatching} and \ref{fig:DegenerateLCs}. This can then be 
used to guide modeling efforts, especially when coupled with other constraints, 
such as a measurement of the core mass and thereby progenitor mass at the time 
of explosion (as in \citealt{Jerkstrand2012}). With a clear independent 
constraint on one explosion parameter, such as an observed progenitor radius, 
the other explosion properties can be simply recovered to around 15\%. 

\acknowledgements
We thank Josiah Schwab for guidance interpreting \STELLA\ output and for 
formative conversations, Evan Bauer for discussions and comments on the 
original draft of the manuscript, and Dan Kasen for extremely helpful discussions. 
We also thank Claudia Gutierrez and Ondrej Pejcha for graciously providing data, 
and Luc Dessart for providing models to compare to. This research benefited 
from interactions with 
Maria Drout,
Paul Duffell, 
Jim Fuller, 
Sterl Phinney,
Eliot Quataert, 
and
Todd Thompson
that were funded by the Gordon and Betty Moore Foundation through Grant GBMF5076. 
It is a pleasure also to thank 
Azalee Boestrom, 
Daichi Hiramatsu,
Viktoriya Morozova,
and
Stefano Valenti
for discussions and correspondences about observations. 
Lastly, we thank the anonymous referee for constructive comments 
that helped improve this manuscript.

J.A.G. is supported by the National Science Foundation Graduate Research Fellowship under 
grant number 1650114. The \MESA\ project is supported by the National Science Foundation (NSF)
under the Software Infrastructure for Sustained Innovation program grant ACI-1663688.
This research was supported in part by the Gordon and Betty Moore Foundation
through Grant GBMF5076 and at the KITP by the NSF under grant PHY-1748958.

This research made extensive use of the SAO/NASA Astrophysics Data System (ADS).

\software{
\texttt{MesaScript} \citep{MesaScript}, 
\texttt{Python} from \href{https://www.python.org}{python.org}, 
\texttt{py\_mesa\_reader} \citep{MesaReader}, 
\texttt{ipython/jupyter} \citep{perez2007ipython,kluyver2016jupyter}, 
\texttt{SciPy} \citep{scipy}, 
\texttt{NumPy} \citep{der_walt_2011_aa}, and 
\texttt{matplotlib} \citep{hunter_2007_aa}.
}

\appendix
\section{Quantifying Fallback in Core-Collapse Supernovae \label{sec:FALLBACK}}

Here we discuss modifications relative to MESA IV, of \MESA\ modeling of the ejecta evolution
after core collapse in massive stars (roughly M $>8 \Msun$). These are focused on cases 
where the total final explosion energy is positive, 
but insufficient to unbind the entirety of the material which does not 
initially collapse into the compact object.
In these weak explosions, there is some amount of fallback
material which does not become unbound.
Our emphasis here is to quantify and remove fallback in model explosions
of RSG progenitor stars. 
Although we describe models of Type IIP SN explosions, this scheme can be similarly applied to core 
collapse events in massive stars which have lost the majority of their outer Hydrogen envelope,
which produce Type IIb and Ib SNe. 

In MESA IV, three options existed to treat fallback: 
\begin{enumerate}
\item Set the velocity of all 
inward-moving material with negative total energy to be zero, which creates a hydrostatic 
shell that can be excised from the ejecta before handing off to the radiation hydrodynamics 
code \STELLA\ to calculate SN observables. 
\item During the 
shock propagation phase, remove material at the inner boundary (IB) if it has 
negative velocity (i.e. if it is infalling). 
\item Remove material at the IB
if it is moving with negative velocity and also has net negative energy 
(i.e. it is bound and infalling). 
\end{enumerate}

However, triggering fallback based only on conditions in the innermost zone can lead to problems. 
For example, in many models at lower explosion energies, while the innermost zone may have 
negative cell-centered velocity, it can be in thermal contact with neighboring zones. Therefore 
to remove cells solely based upon their having negative velocity creates a vacuum at the 
IB which can remove energy and mass which could otherwise remain in the ejecta. Moreover, 
energy deposited at the IB by any inward-propagating shock
can cause the innermost zones to have positive total energy, 
while being surrounded by a larger amount of material with net negative energy. 
Because of this, in some models, checking only if the innermost zone is bound before triggering 
fallback can lead to bound material piling up on top of a small number of cells with positive 
total energy. If not removed this can lead to a globally bound hydrostatic shell building 
up in the center, which might interact with the ejecta and affect concentrations 
of important species such as H and $\Ni$, thus affecting SN properties.  Such a region can 
also lead to numerical problems if not properly excised before handing off to radiative transfer
codes such as \STELLA.

\begin{figure}
\centering
\includegraphics[width=0.45\columnwidth]{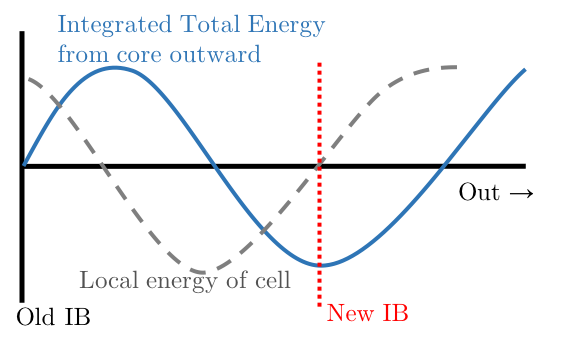} 
\caption{Diagram for new fallback criterion at a timestep 
where there is net positive energy near the inner boundary, but a larger, 
gravitationally bound region above it which will eventually fall back.}
\label{f.fallbackpng}
\end{figure}

\citet{Paxton2019} (MESA 5) introduces two new user controls to better account for material which could fall 
onto the central object during the hydrodynamical evolution of low explosion energy
core-collapse SNe. First, a new criterion is implemented to select which material is excised 
from the model.\footnote{This criterion is triggered when \texttt{fallback\_check\_total\_energy} is 
set to \texttt{.true.} in \texttt{star\_job}.} At each timestep, \MESA\ calculates the 
integrated total energy from the innermost cell to cell $j$ above it:
\begin{equation}
E_j = \sum_{i={\rm inner}}^{j} \left[e_i - \frac{G m_i}{r_i} + \frac{1}{2} u_i^2 \right]{\rm d}m_i,
\label{e.energyintegral}
\end{equation} 
where for cell $i$ at mass $m_i$ and radius $r_i$, $e_i$ is the internal energy in erg g$^{-1}$ and 
$u_i$ is the velocity in cm s$^{-1}$. If $E_j < 0$, then there is a bound inner region, and \MESA\ 
continues this calculation outward until it reaches a cell $k$ with local positive total energy 
($e_k - G m_k/r_k + u_k^2/2>0$), causing the integral to be at a local minimum. 
\MESA\ deletes material inside this zone, and moves the IB, fixing the inner radius of zone 
$k$ to be the new radius of the inner boundary \verb|r_center|, and setting the velocity 
at the inner boundary \verb|v_center|=0. A schematic diagram of this calculation, 
in a case where fallback is triggered but the innermost zones are unbound, 
is shown in Figure \ref{f.fallbackpng}. 

\begin{figure}
\centering
\includegraphics{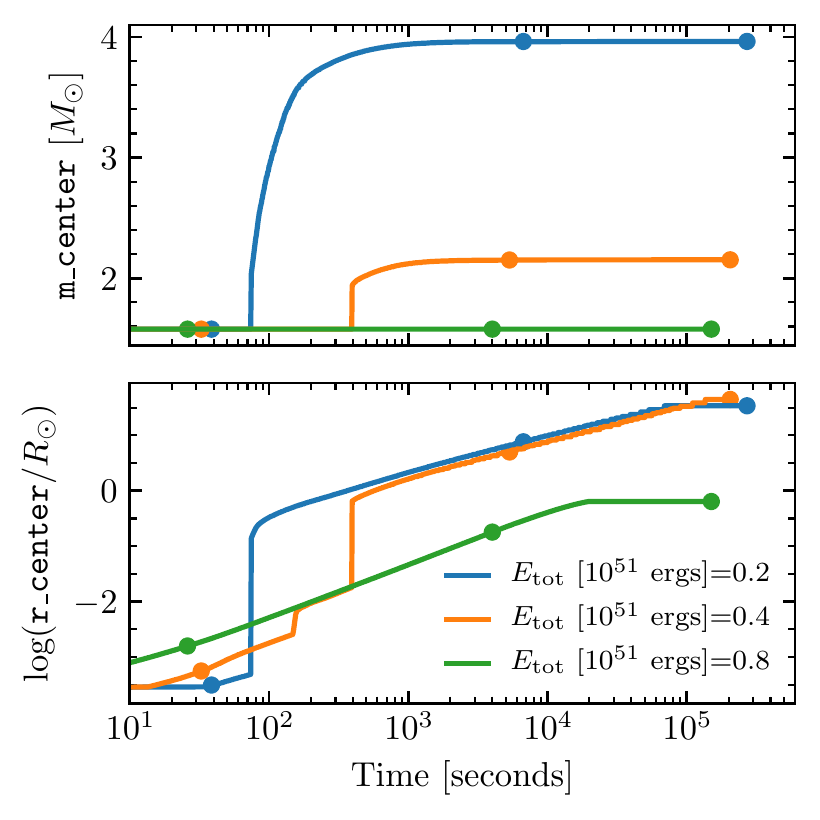} 
\caption{Evolution of the mass (upper panel) and radial (lower panel) coordinate of the inner boundary
for the new fallback prescription for the M12.9\_R766 progenitor model from \mesafour\ 
for explosions of increasing energy. Colored dots correspond to times that the forward shock 
crosses a sharp compositional gradient: entering the He layer, entering the H layer, and 
shock breakout. Because the fallback prescription holds \texttt{r\_center}
fixed and \texttt{v\_center}$=0$, all changes in \texttt{r\_center} result from inner cells 
being removed from the model.}
\label{f.fallbackcenter}
\end{figure}

Figure \ref{f.fallbackcenter} shows the evolution of the inner boundary for explosions 
of varying total energy just after the explosion ($\Etot$, defined in Section \ref{sec:MODELS}),
using the new fallback criterion for the M12.9\_R766 progenitor model, 
which has a total energy of $-4.4\times10^{50}$ ergs just before the 
explosion. Nearly all of the mass lost to fallback occurs while the forward-moving shock 
is in the Helium layer, beginning around the time that the reverse shock generated at 
the interface between the CO/He layers reaches the inner boundary. Because the new fallback 
prescription sets \verb|v_center|=0 and fixes \verb|r_center| except in the case of 
fallback being triggered, all changes in the radius of the inner boundary are due to 
cells being removed from the inner boundary. For sufficiently large explosion energies, 
little to no fallback is seen, although some cells of negligible mass are removed from 
the inner boundary, causing the radius of the inner boundary to move outward. 

Second, in order to remove any slow-moving, nearly hydrostatic material left 
near the inner boundary as a result of the fixed \texttt{r\_center}, which 
may cause problems after handing off to radiation hydrodynamic codes 
(see Figure \ref{f.fallbackvcutlight curves}), \MESA\ allows the user to specify 
a minimum innermost velocity for material which gets included in the final 
ejecta profile that is handed off to \STELLA.\footnote{This is controlled by 
the\texttt{star\_job} inlist parameter \texttt{stella\_skip\_inner\_v\_limit}, which is the minimum 
velocity of the inner ejecta to include in the profile handed off to \STELLA\ in units of 
cm s$^{-1}$. } \MESA\ will then exclude all material beneath the innermost zone that has 
velocity greater than this velocity cut. This can lead to a small amount of additional 
mass which is excluded from the final ejecta profile at handoff. 

\begin{figure}
\centering
\includegraphics{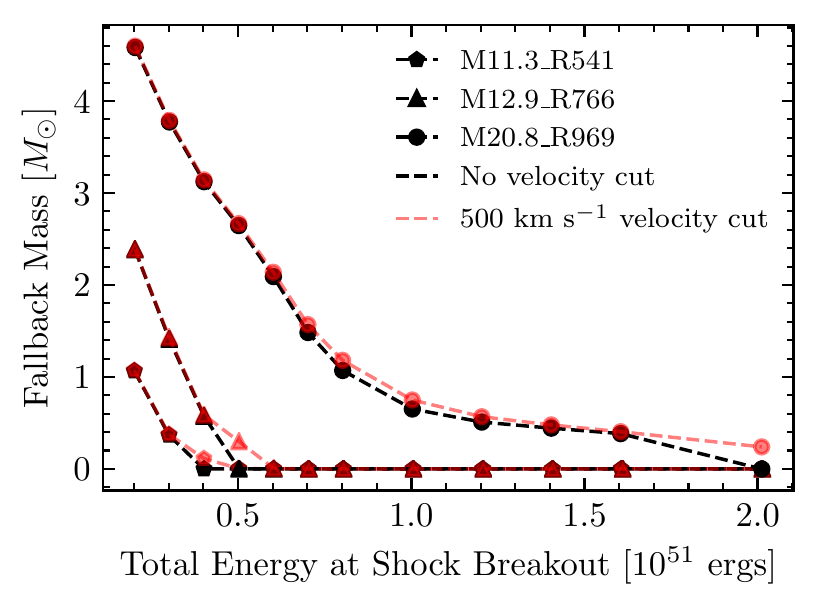}
\caption{Mass of fallback material at shock breakout as a function of the total energy 
of the exploded model at the time of shock breakout for three different progenitor models 
exploded with 12 different explosion energies. 
Results are shown for the new integrated energy fallback criterion with no additional 
velocity cut (black points), and the same criterion with a 500~km~s$^{-1}$ velocity cut 
at shock breakout (red points).}
\label{f.fallbackoutcome}
\end{figure}

\begin{figure}
\centering
\includegraphics{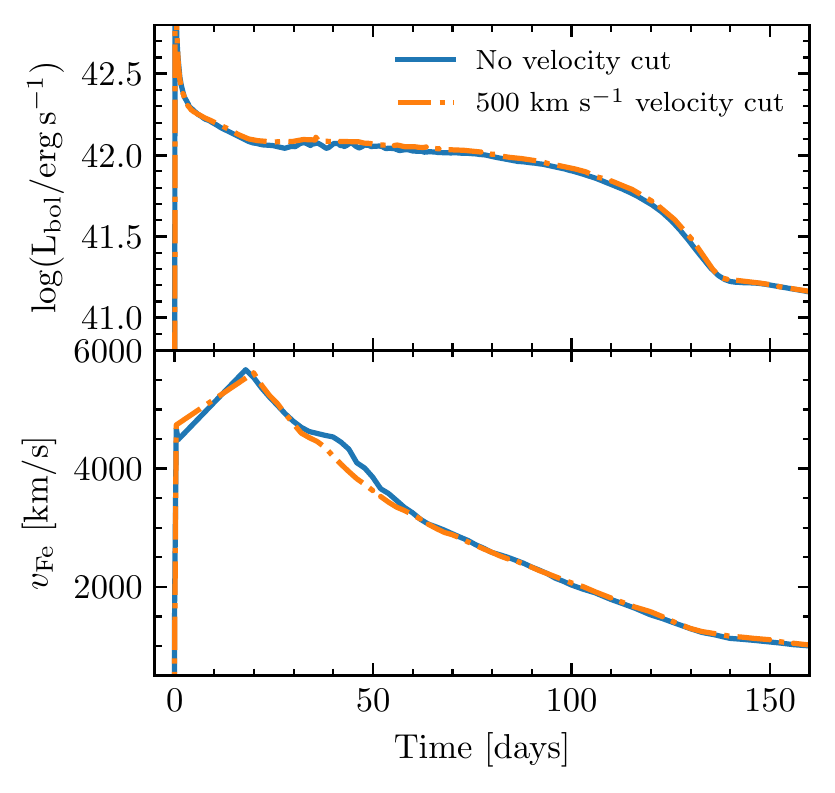} 
\caption{Effects of a velocity cut on \STELLA\ light curves and Fe II 5169\AA\ line velocities 
for our M12.9\_R766 progenitor model exploded with $E_{\rm tot} = 5.0\times 10^{51}$ ergs 
and a nickel mass $M_{\rm Ni}=0.042\Msun$, where we see a noticeable difference between 
the mass of fallback material with and without a velocity cut ($\approx0.3\Msun$).}
\label{f.fallbackvcutlight curves}
\end{figure}

The result of both modifications is shown in Figure \ref{f.fallbackoutcome}, for three different 
models exploded at 12 different explosion energies. This can be loosely compared with 
Figure 6 of \citet{Perna2014}.
Included are the M12.9\_R766 and M11.3\_R541 models 
from our standard suite, as well as an additional model, named M20.8\_R969, which has
binding energy $-8.4\times10^{50}$ ergs just before the explosion, included in order 
to demonstrate an explosion in a more massive star where there would be more fallback 
material due to more strongly bound core material. Generally, models with and without a
velocity cut end with roughly the same amount of fallback. In cases where 
the explosion energy is just barely enough to unbind all of the 
mass, the velocity cut can remove a small additional amount of material. However, 
as seen in Figure \ref{f.fallbackvcutlight curves}, even in this case, a suitable 
velocity cut between 100 - 500 km s$^{-1}$ has very little effect on light curve properties 
and the photospheric evolution of the SN, and can greatly reduce numerical artifacts which 
may arise from an inward-propagating shock hitting the inner boundary in \STELLA. 
Such a cut also can lead to a factor of 10 or more speedup in number of timesteps required 
to produce a light curve.

\section{Extension of the Plateau due to $\Ni$ Decay\label{sec:IIp_equations}}

We start with the thermodynamic equation, where a fluid is heated by nuclear decay (in our case, of $\Ni$)
with complete trapping

\begin{equation}
TdS = dE + PdV = L_{\rm nuc}dt
\end{equation}

In a 1-zone, radiation-dominated regime, we can express $P = E/3V$ and $V = 4\pi (v t)^3/3$. 
Assuming homology, $dV/V = 3dt/t$, and this becomes 
\begin{equation}
\frac{1}{t}d(Et) = L_{\rm nuc}dt.
\end{equation}

To find the total energy at time $t$, integrate from from $\tSB$ to obtain:
\begin{equation}
E(t) =  E_0 \frac{\tSB}{t} + \frac{1}{t} \int_{\tSB}^{t} t' L_{\rm nuc}\,dt',
\label{eq:useful}
\end{equation}
where $L_{\rm nuc}$ is due to the $\Ni\rightarrow \Co\rightarrow \Fe$ 
decay chain, following \cite{Nadyozhin1994}:
\begin{equation}
L_{\rm nuc} = \frac{\NNi Q_{\rm Ni} }{\tau_{\rm Ni}} \exp\left(\frac{-t}{\tau_{\rm Ni}}\right) 
+ \frac{ \NNi Q_{\rm Co}}{\tau_{\rm Co} - \tau_{\rm Ni}}
\left[\exp\left(\frac{-t}{\tau_{\rm Co}}\right) 
    - \exp\left(\frac{-t}{\tau_{\rm Ni}}\right)\right],
\end{equation}

where $\NNi = M_{{\rm Ni}}/(56\, {\rm amu})$, $\tau_{\rm X}$ is the lifetime of radioactive species X, 
and $Q_{\rm X}$ is the energy per decay of species X.

Assuming only $\Ni$ is produced in the explosion and all $\Co$ comes from $\Ni$ decay,
the contribution to the internal energy due to the $\Ni$ decay chain over the lifetime of the SN is 
\begin{equation}
E_{\rm tot, Ni} =\frac{1}{\tpt} \int_{\tSB}^{\tpt} t L_{\rm nuc}\,dt .
\end{equation}

We now make a few approximations: First, by the end of the plateau, 
$\Ni$ has undergone many decay times. Thus we take $\tSB \rightarrow 0$ and 
$\tpt/\tau_{\rm Ni} \rightarrow \infty$ when in the bounds of our integrals.
However, the decay time of $^{56}$Co is 111.3 days, which is comparable to $\tpt$. Thus we 
approximate $\tpt/\tau_{\rm Co} \approx 1$ when in the bounds of our integrals. 
Outside the integrals, we assume that the time to shock breakout is roughly the expansion time, 
$\tsb\approx\te$, where, as in Section \ref{sec:PEDAGOGY},
\begin{align*}
\te &= R_0/v_{\rm e},\\
v_{\rm e} &= \sqrt{\frac{2\Eexp}{\Mej}} \approx 3.16\times10^8 M_{10}E_{51}\mathrm{\ cm\ s^{-1}}.
\end{align*}
Any numerical quantities are, in reality, dependent on the specifics of the relevant timescales. 
Here we aim primarily to capture the relevant scaling relationships, fitting against 
our models to find appropriate numerical prefactors.

Computing these integrals and simplifying, we find that 
\begin{align}
E_{\rm int}(\tpt) &= \frac{E_0\, t_e}{\tpt} + \frac{\NNi}{\tpt} \left[Q_{\rm Ni} \tau_{\rm Ni}
+ Q_{\rm Co} \left(\frac{0.26 \tau_{\rm Co}^2 - \tau_{\rm Ni}^2}{\tau_{\rm Co} - \tau_{\rm Ni}}\right)
\right]\\
&= \frac{E_0\, t_e}{\tpt} \times f_{\rm rad},\ {\rm where}\\
f_{\rm rad} &\equiv 1 + \frac{\NNi}{t_e E_0}\left(Q_{\rm Ni} \tau_{\rm Ni} + Q_{\rm Co} \tau'_{\rm Co}\right)\ {\rm and}\\
\tau'_{\rm Co} &\equiv  \left(\frac{0.26 \tau_{\rm Co}^2 - \tau_{\rm Ni}^2}{\tau_{\rm Co} - \tau_{\rm Ni}}\right),
\end{align}
noting that $E_0,$ the internal energy at $\tSB$, is roughly half the total energy of the explosion 
(mentioned as a comment in K\&W), we set $E_0 = \Eexp/2$.

We can re-express $f_{\rm rad}$ as 
\begin{equation}
f_{\rm rad} = 1 + \frac{\MNi}{E_0}\frac{\tau_{\rm Ni}}{t_e}\left(q_{\rm Ni} + q_{\rm Co} \frac{\tau'_{\rm Co}}{\tau_{\rm Ni}}\right)
\end{equation}
where $q_{\rm X}$ is the specific (per gram) energy released by the decay of species X; in this case 
$q_{\rm X} = Q_{\rm X}/56\, {\rm amu}$.

Following \citet{Nadyozhin1994}, we use $Q_{\rm Ni} = 1.75$ MeV/nucleon, 
$Q_{\rm Co} = 3.73$ MeV/nucleon, $\tau_{\rm Ni} = 8.8$ days, and $\tau_{\rm Co} = 111.3$ days. We thus find that  
\begin{equation}
f_{\rm rad} \approx 1 + 7.0 \left(M_{{\rm Ni}, \odot}\, \Efoe^{-1/2}\, 
R_{500}^{-1}\,\Mten^{-1/2} \right).
\label{eq:frad_scaling}
\end{equation}

This argument ignores the effects of the distribution of $\Ni$, as we 
necessarily have assumed in this simple 1-zone model that the nickel is distributed evenly 
throughout the ejecta. If the heat from the $\Ni$ decay is trapped inside the core of the star until 
that material becomes optically thin, then this would further extend the duration of the plateau. 
Thus, we should treat the factor of 7.0 as a rough lower bound, rather than an expectation. 

We can also recast Equation \eqref{eq:frad_scaling} in terms of $\ET$ and $\etaNi$. 
Although our derivation assumes all internal energy is trapped to be radiated away, and the 
\citet{Shussman2016} derivation of $\ET$ assumes that all energy is radiated away, this is 
just a difference in terms and not a difference in physics. 
Thus at $t=\tpt$, plugging in $\ET = E_0 \tSB \approx E_0 t_e$ and $\etaNi= \left(\int_{\tSB}^{\tpt} t L_{\rm nuc}\,dt\right)/\ET$ 
to Equation \eqref{eq:useful}, we recover 
\begin{equation}
E_{\rm int}(\tpt) =  E_0 \frac{t_e}{\tpt} + \frac{\ET\, \etaNi}{\tpt} \approx \frac{E_0\, t_e}{\tpt}(1 + \etaNi),
\end{equation}
so $f_{\rm rad} \approx  1 + \etaNi$.

\bibliographystyle{yahapj}
\singlespace
\bibliography{manuscript}

\begin{thebibliography}{}
\providecommand\natexlab[1]{#1}
\providecommand\JournalTitle[1]{#1}

\bibitem[{{Arnett}(1980)}]{Arnett1980}
{Arnett}, W.~D. 1980,
  \href{http://dx.doi.org/10.1086/157898}{\JournalTitle{\apj}, 237, 541}

\bibitem[{{Baklanov} {et~al.}(2005){Baklanov}, {Blinnikov}, \&
  {Pavlyuk}}]{Baklanov2005}
{Baklanov}, P.~V., {Blinnikov}, S.~I., \& {Pavlyuk}, N.~N. 2005,
  \href{http://dx.doi.org/10.1134/1.1958107}{\JournalTitle{Astronomy Letters},
  31, 429}

\bibitem[{{Bellm} {et~al.}(2019){Bellm}, {Kulkarni}, {Graham}, {Dekany},
  {Smith}, {Riddle}, {Masci}, {Helou}, {Prince}, {Adams}, {Barbarino},
  {Barlow}, {Bauer}, {Beck}, {Belicki}, {Biswas}, {Blagorodnova}, {Bodewits},
  {Bolin}, {Brinnel}, {Brooke}, {Bue}, {Bulla}, {Burruss}, {Cenko}, {Chang},
  {Connolly}, {Coughlin}, {Cromer}, {Cunningham}, {De}, {Delacroix}, {Desai},
  {Duev}, {Eadie}, {Farnham}, {Feeney}, {Feindt}, {Flynn}, {Franckowiak},
  {Frederick}, {Fremling}, {Gal-Yam}, {Gezari}, {Giomi}, {Goldstein},
  {Golkhou}, {Goobar}, {Groom}, {Hacopians}, {Hale}, {Henning}, {Ho}, {Hover},
  {Howell}, {Hung}, {Huppenkothen}, {Imel}, {Ip}, {Ivezi{\'c}}, {Jackson},
  {Jones}, {Juric}, {Kasliwal}, {Kaspi}, {Kaye}, {Kelley}, {Kowalski},
  {Kramer}, {Kupfer}, {Landry}, {Laher}, {Lee}, {Lin}, {Lin}, {Lunnan},
  {Giomi}, {Mahabal}, {Mao}, {Miller}, {Monkewitz}, {Murphy}, {Ngeow},
  {Nordin}, {Nugent}, {Ofek}, {Patterson}, {Penprase}, {Porter}, {Rauch},
  {Rebbapragada}, {Reiley}, {Rigault}, {Rodriguez}, {van Roestel}, {Rusholme},
  {van Santen}, {Schulze}, {Shupe}, {Singer}, {Soumagnac}, {Stein}, {Surace},
  {Sollerman}, {Szkody}, {Taddia}, {Terek}, {Van Sistine}, {van Velzen},
  {Vestrand}, {Walters}, {Ward}, {Ye}, {Yu}, {Yan}, \& {Zolkower}}]{ZTF2019}
{Bellm}, E.~C., {Kulkarni}, S.~R., {Graham}, M.~J., {et~al.} 2019,
  \href{http://dx.doi.org/10.1088/1538-3873/aaecbe}{\JournalTitle{\pasp}, 131,
  018002}

\bibitem[{Bersten \& Hamuy(2009)}]{Bersten2009}
Bersten, M.~C., \& Hamuy, M. 2009,
  \href{http://dx.doi.org/10.1088/0004-637x/701/1/200}{\JournalTitle{\apj},
  701, 200}

\bibitem[{{Blinnikov} \& {Sorokina}(2004)}]{Blinnikov2004}
{Blinnikov}, S., \& {Sorokina}, E. 2004,
  \href{http://dx.doi.org/10.1023/B:ASTR.0000022161.03559.42}{\JournalTitle{\apss},
  290, 13}

\bibitem[{{Blinnikov} {et~al.}(1998){Blinnikov}, {Eastman}, {Bartunov},
  {Popolitov}, \& {Woosley}}]{Blinnikov1998}
{Blinnikov}, S.~I., {Eastman}, R., {Bartunov}, O.~S., {Popolitov}, V.~A., \&
  {Woosley}, S.~E. 1998,
  \href{http://dx.doi.org/10.1086/305375}{\JournalTitle{\apj}, 496, 454}

\bibitem[{{Blinnikov} {et~al.}(2006){Blinnikov}, {R{\"o}pke}, {Sorokina},
  {Gieseler}, {Reinecke}, {Travaglio}, {Hillebrandt}, \&
  {Stritzinger}}]{Blinnikov2006}
{Blinnikov}, S.~I., {R{\"o}pke}, F.~K., {Sorokina}, E.~I., {et~al.} 2006,
  \href{http://dx.doi.org/10.1051/0004-6361:20054594}{\JournalTitle{\aap}, 453,
  229}

\bibitem[{Brayton {et~al.}(1972)Brayton, Gustavson, \& Hachtel}]{Brayton1972}
Brayton, R.~K., Gustavson, F.~G., \& Hachtel, G.~D. 1972,
  \href{http://dx.doi.org/10.1109/PROC.1972.8562}{\JournalTitle{Proceedings of
  the IEEE}, 60, 98}

\bibitem[{{Brown} {et~al.}(2013){Brown}, {Baliber}, {Bianco}, {Bowman},
  {Burleson}, {Conway}, {Crellin}, {Depagne}, {De Vera}, {Dilday}, {Dragomir},
  {Dubberley}, {Eastman}, {Elphick}, {Falarski}, {Foale}, {Ford}, {Fulton},
  {Garza}, {Gomez}, {Graham}, {Greene}, {Haldeman}, {Hawkins}, {Haworth},
  {Haynes}, {Hidas}, {Hjelstrom}, {Howell}, {Hygelund}, {Lister}, {Lobdill},
  {Martinez}, {Mullins}, {Norbury}, {Parrent}, {Paulson}, {Petry}, {Pickles},
  {Posner}, {Rosing}, {Ross}, {Sand}, {Saunders}, {Shobbrook}, {Shporer},
  {Street}, {Thomas}, {Tsapras}, {Tufts}, {Valenti}, {Vander Horst}, {Walker},
  {White}, \& {Willis}}]{LCO2013}
{Brown}, T.~M., {Baliber}, N., {Bianco}, F.~B., {et~al.} 2013,
  \href{http://dx.doi.org/10.1086/673168}{\JournalTitle{\pasp}, 125, 1031}

\bibitem[{{Burrows} {et~al.}(2019){Burrows}, {Radice}, \&
  {Vartanyan}}]{Burrows2019}
{Burrows}, A., {Radice}, D., \& {Vartanyan}, D. 2019,
  \href{http://dx.doi.org/10.1093/mnras/stz543}{\JournalTitle{\mnras}, 538}

\bibitem[{{Castor}(1970)}]{Castor1970}
{Castor}, J.~I. 1970,
  \href{http://dx.doi.org/10.1093/mnras/149.2.111}{\JournalTitle{\mnras}, 149,
  111}

\bibitem[{{Chugai}(1991)}]{Chugai1991}
{Chugai}, N.~N. 1991, \JournalTitle{Soviet Astronomy Letters}, 17, 210

\bibitem[{{Dessart} \& {Hillier}(2019)}]{Dessart2019}
{Dessart}, L., \& {Hillier}, D.~J. 2019, \JournalTitle{arXiv e-prints},
  arXiv:1903.04840

\bibitem[{{Dessart} {et~al.}(2013){Dessart}, {Hillier}, {Waldman}, \&
  {Livne}}]{Dessart2013}
{Dessart}, L., {Hillier}, D.~J., {Waldman}, R., \& {Livne}, E. 2013,
  \href{http://dx.doi.org/10.1093/mnras/stt861}{\JournalTitle{\mnras}, 433,
  1745}

\bibitem[{{Duffell}(2016)}]{Duffell2016}
{Duffell}, P.~C. 2016,
  \href{http://dx.doi.org/10.3847/0004-637X/821/2/76}{\JournalTitle{\apj}, 821,
  76}

\bibitem[{{Gear}(1971)}]{Gear1971}
{Gear}, C.~W. 1971, {Numerical initial value problems in ordinary differential
  equations}, {Prentice-Hall Series in Automatic Computation} ({Prentice-Hall,
  Engelwood Cliffs})

\bibitem[{Guti{\'{e}}rrez {et~al.}(2017)Guti{\'{e}}rrez, Anderson, Hamuy,
  Gonz{\'{a}}lez-Gaitan, Galbany, Dessart, Stritzinger, Phillips, Morrell, \&
  Folatelli}]{Gutierrez2017}
Guti{\'{e}}rrez, C.~P., Anderson, J.~P., Hamuy, M., {et~al.} 2017,
  \href{http://dx.doi.org/10.3847/1538-4357/aa8f42}{\JournalTitle{\apj}, 850,
  90}

\bibitem[{Hamuy(2003)}]{Hamuy2003}
Hamuy, M. 2003, \href{http://dx.doi.org/10.1086/344689}{\JournalTitle{\apj},
  582, 905}

\bibitem[{Hunter(2007)}]{hunter_2007_aa}
Hunter, J.~D. 2007, \JournalTitle{Computing In Science \&amp; Engineering}, 9,
  90

\bibitem[{{Jerkstrand} {et~al.}(2012){Jerkstrand}, {Fransson}, {Maguire},
  {Smartt}, {Ergon}, \& {Spyromilio}}]{Jerkstrand2012}
{Jerkstrand}, A., {Fransson}, C., {Maguire}, K., {et~al.} 2012,
  \href{http://dx.doi.org/10.1051/0004-6361/201219528}{\JournalTitle{\aap},
  546, A28}

\bibitem[{Jones {et~al.}(2001--)Jones, Oliphant, Peterson, {et~al.}}]{scipy}
Jones, E., Oliphant, T., Peterson, P., {et~al.} 2001--, {SciPy}: Open source
  scientific tools for {Python}

\bibitem[{{Kasen} {et~al.}(2006){Kasen}, {Thomas}, \& {Nugent}}]{Kasen2006}
{Kasen}, D., {Thomas}, R.~C., \& {Nugent}, P. 2006,
  \href{http://dx.doi.org/10.1086/506190}{\JournalTitle{\apj}, 651, 366}

\bibitem[{{Kasen} \& {Woosley}(2009)}]{Kasen2009}
{Kasen}, D., \& {Woosley}, S.~E. 2009,
  \href{http://dx.doi.org/10.1088/0004-637X/703/2/2205}{\JournalTitle{\apj},
  703, 2205}

\bibitem[{Kluyver {et~al.}(2016)Kluyver, Ragan-Kelley, P{\'e}rez, Granger,
  Bussonnier, Frederic, Kelley, Hamrick, Grout, Corlay,
  {et~al.}}]{kluyver2016jupyter}
Kluyver, T., Ragan-Kelley, B., P{\'e}rez, F., {et~al.} 2016, in Positioning and
  Power in Academic Publishing: Players, Agents and Agendas: Proceedings of the
  20th International Conference on Electronic Publishing, IOS Press, 87

\bibitem[{{Kochanek} {et~al.}(2017){Kochanek}, {Shappee}, {Stanek}, {Holoien},
  {Thompson}, {Prieto}, {Dong}, {Shields}, {Will}, {Britt}, {Perzanowski}, \&
  {Pojma{\'n}ski}}]{ASSASN2017}
{Kochanek}, C.~S., {Shappee}, B.~J., {Stanek}, K.~Z., {et~al.} 2017,
  \href{http://dx.doi.org/10.1088/1538-3873/aa80d9}{\JournalTitle{Publications
  of the Astronomical Society of the Pacific}, 129, 104502}

\bibitem[{Kozyreva {et~al.}(2018)Kozyreva, Nakar, \& Waldman}]{Kozyreva2018}
Kozyreva, A., Nakar, E., \& Waldman, R. 2018,
  \href{http://dx.doi.org/10.1093/mnras/sty3185}{\JournalTitle{\mnras}, 483,
  1211}

\bibitem[{{Lisakov} {et~al.}(2017){Lisakov}, {Dessart}, {Hillier}, {Waldman},
  \& {Livne}}]{Lisakov2017}
{Lisakov}, S.~M., {Dessart}, L., {Hillier}, D.~J., {Waldman}, R., \& {Livne},
  E. 2017,
  \href{http://dx.doi.org/10.1093/mnras/stw3035}{\JournalTitle{\mnras}, 466,
  34}

\bibitem[{Litvinova \& Nadyozhin(1983)}]{Litvinova1983}
Litvinova, I.~Y., \& Nadyozhin, D.~K. 1983,
  \href{http://dx.doi.org/10.1007/BF01008387}{\JournalTitle{Astrophysics and
  Space Science}, 89, 89}

\bibitem[{{LSST Science Collaboration} {et~al.}(2009){LSST Science
  Collaboration}, {Abell}, {Allison}, {Anderson}, {Andrew}, {Angel}, {Armus},
  {Arnett}, {Asztalos}, {Axelrod}, \& et~al.}]{LSST2009}
{LSST Science Collaboration}, {Abell}, P.~A., {Allison}, J., {et~al.} 2009,
  \JournalTitle{arXiv e-prints},
  \href{http://arxiv.org/abs/0912.0201}{{\sffamily arXiv:0912.0201
  [astro-ph.IM]}}

\bibitem[{{Mihalas}(1978)}]{Mihalas1978}
{Mihalas}, D. 1978, {Stellar atmospheres /2nd edition/}

\bibitem[{{Morozova} {et~al.}(2016){Morozova}, {Piro}, {Renzo}, \&
  {Ott}}]{Morozova2016}
{Morozova}, V., {Piro}, A.~L., {Renzo}, M., \& {Ott}, C.~D. 2016,
  \href{http://dx.doi.org/10.3847/0004-637X/829/2/109}{\JournalTitle{\apj},
  829, 109}

\bibitem[{{Morozova} {et~al.}(2017){Morozova}, {Piro}, \&
  {Valenti}}]{Morozova2017}
{Morozova}, V., {Piro}, A.~L., \& {Valenti}, S. 2017,
  \href{http://dx.doi.org/10.3847/1538-4357/aa6251}{\JournalTitle{\apj}, 838,
  28}

\bibitem[{{M{\"u}ller} {et~al.}(2017){M{\"u}ller}, {Prieto}, {Pejcha}, \&
  {Clocchiatti}}]{Muller2017}
{M{\"u}ller}, T., {Prieto}, J.~L., {Pejcha}, O., \& {Clocchiatti}, A. 2017,
  \href{http://dx.doi.org/10.3847/1538-4357/aa72f1}{\JournalTitle{\apj}, 841,
  127}

\bibitem[{{Nadyozhin}(1994)}]{Nadyozhin1994}
{Nadyozhin}, D.~K. 1994,
  \href{http://dx.doi.org/10.1086/192008}{\JournalTitle{\apjs}, 92, 527}

\bibitem[{{Nakar} {et~al.}(2016){Nakar}, {Poznanski}, \& {Katz}}]{Nakar2016}
{Nakar}, E., {Poznanski}, D., \& {Katz}, B. 2016,
  \href{http://dx.doi.org/10.3847/0004-637X/823/2/127}{\JournalTitle{\apj},
  823, 127}

\bibitem[{{Paxton} {et~al.}(2011){Paxton}, {Bildsten}, {Dotter}, {Herwig},
  {Lesaffre}, \& {Timmes}}]{Paxton2011}
{Paxton}, B., {Bildsten}, L., {Dotter}, A., {et~al.} 2011,
  \href{http://dx.doi.org/10.1088/0067-0049/192/1/3}{\JournalTitle{\apjs}, 192,
  3}

\bibitem[{{Paxton} {et~al.}(2013){Paxton}, {Cantiello}, {Arras}, {Bildsten},
  {Brown}, {Dotter}, {Mankovich}, {Montgomery}, {Stello}, {Timmes}, \&
  {Townsend}}]{Paxton2013}
{Paxton}, B., {Cantiello}, M., {Arras}, P., {et~al.} 2013,
  \href{http://dx.doi.org/10.1088/0067-0049/208/1/4}{\JournalTitle{\apjs}, 208,
  4}

\bibitem[{{Paxton} {et~al.}(2015){Paxton}, {Marchant}, {Schwab}, {Bauer},
  {Bildsten}, {Cantiello}, {Dessart}, {Farmer}, {Hu}, {Langer}, {Townsend},
  {Townsley}, \& {Timmes}}]{Paxton2015}
{Paxton}, B., {Marchant}, P., {Schwab}, J., {et~al.} 2015,
  \href{http://dx.doi.org/10.1088/0067-0049/220/1/15}{\JournalTitle{\apjs},
  220, 15}

\bibitem[{{Paxton} {et~al.}(2018){Paxton}, {Schwab}, {Bauer}, {Bildsten},
  {Blinnikov}, {Duffell}, {Farmer}, {Goldberg}, {Marchant}, {Sorokina},
  {Thoul}, {Townsend}, \& {Timmes}}]{Paxton2018}
{Paxton}, B., {Schwab}, J., {Bauer}, E.~B., {et~al.} 2018,
  \href{http://dx.doi.org/10.3847/1538-4365/aaa5a8}{\JournalTitle{\apjs}, 234,
  34}

\bibitem[{{Paxton} {et~al.}(2019){Paxton}, {Smolec}, {Gautschy}, {Bildsten},
  {Cantiello}, {Dotter}, {Farmer}, {Goldberg}, {Jermyn}, {Kanbur}, {Marchant},
  {Schwab}, {Thoul}, {Townsend}, {Wolf}, {Zhang}, \& {Timmes}}]{Paxton2019}
{Paxton}, B., {Smolec}, R., {Gautschy}, A., {et~al.} 2019, \JournalTitle{arXiv
  e-prints}, arXiv:1903.01426

\bibitem[{{Pejcha} \& {Prieto}(2015{\natexlab{a}})}]{Pejcha2015a}
{Pejcha}, O., \& {Prieto}, J.~L. 2015{\natexlab{a}},
  \href{http://dx.doi.org/10.1088/0004-637X/799/2/215}{\JournalTitle{\apj},
  799, 215}

\bibitem[{{Pejcha} \& {Prieto}(2015{\natexlab{b}})}]{Pejcha2015b}
---. 2015{\natexlab{b}},
  \href{http://dx.doi.org/10.1088/0004-637X/806/2/225}{\JournalTitle{\apj},
  806, 225}

\bibitem[{P{\'e}rez \& Granger(2007)}]{perez2007ipython}
P{\'e}rez, F., \& Granger, B.~E. 2007, \JournalTitle{Computing in Science \&
  Engineering}, 9, 21

\bibitem[{{Perna} {et~al.}(2014){Perna}, {Duffell}, {Cantiello}, \&
  {MacFadyen}}]{Perna2014}
{Perna}, R., {Duffell}, P., {Cantiello}, M., \& {MacFadyen}, A.~I. 2014,
  \href{http://dx.doi.org/10.1088/0004-637X/781/2/119}{\JournalTitle{\apj},
  781, 119}

\bibitem[{{Popov}(1993)}]{Popov1993}
{Popov}, D.~V. 1993,
  \href{http://dx.doi.org/10.1086/173117}{\JournalTitle{\apj}, 414, 712}

\bibitem[{{Shussman} {et~al.}(2016{\natexlab{a}}){Shussman}, {Nakar},
  {Waldman}, \& {Katz}}]{Shussman2016}
{Shussman}, T., {Nakar}, E., {Waldman}, R., \& {Katz}, B. 2016{\natexlab{a}},
  \JournalTitle{arXiv e-prints},
  \href{http://arxiv.org/abs/1602.02774}{{\sffamily arXiv:1602.02774
  [astro-ph.HE]}}

\bibitem[{{Shussman} {et~al.}(2016{\natexlab{b}}){Shussman}, {Waldman}, \&
  {Nakar}}]{Shussman2016b}
{Shussman}, T., {Waldman}, R., \& {Nakar}, E. 2016{\natexlab{b}},
  \JournalTitle{arXiv e-prints},
  \href{http://arxiv.org/abs/1610.05323}{{\sffamily arXiv:1610.05323
  [astro-ph.HE]}}

\bibitem[{{Smartt}(2009)}]{Smartt2009}
{Smartt}, S.~J. 2009,
  \href{http://dx.doi.org/10.1146/annurev-astro-082708-101737}{\JournalTitle{\araa},
  47, 63}

\bibitem[{{Smartt}(2015)}]{Smartt2015}
---. 2015,
  \href{http://dx.doi.org/10.1017/pasa.2015.17}{\JournalTitle{Publications of
  the Astronomical Society of Australia}, 32, e016}

\bibitem[{{Sobolev}(1960)}]{Sobolev1960}
{Sobolev}, V.~V. 1960, {Moving envelopes of stars}

\bibitem[{{Sukhbold} {et~al.}(2016){Sukhbold}, {Ertl}, {Woosley}, {Brown}, \&
  {Janka}}]{Sukhbold2016}
{Sukhbold}, T., {Ertl}, T., {Woosley}, S.~E., {Brown}, J.~M., \& {Janka}, H.-T.
  2016,
  \href{http://dx.doi.org/10.3847/0004-637X/821/1/38}{\JournalTitle{\apj}, 821,
  38}

\bibitem[{{Utrobin}(2007)}]{Utrobin2007}
{Utrobin}, V.~P. 2007,
  \href{http://dx.doi.org/10.1051/0004-6361:20066078}{\JournalTitle{\aap}, 461,
  233}

\bibitem[{{Utrobin} {et~al.}(2017){Utrobin}, {Wongwathanarat}, {Janka}, \&
  {M{\"u}ller}}]{Utrobin2017}
{Utrobin}, V.~P., {Wongwathanarat}, A., {Janka}, H.-T., \& {M{\"u}ller}, E.
  2017, \href{http://dx.doi.org/10.3847/1538-4357/aa8594}{\JournalTitle{\apj},
  846, 37}

\bibitem[{{Valenti} {et~al.}(2016){Valenti}, {Howell}, {Stritzinger}, {Graham},
  {Hosseinzadeh}, {Arcavi}, {Bildsten}, {Jerkstrand}, {McCully}, {Pastorello},
  {Piro}, {Sand}, {Smartt}, {Terreran}, {Baltay}, {Benetti}, {Brown},
  {Filippenko}, {Fraser}, {Rabinowitz}, {Sullivan}, \& {Yuan}}]{Valenti2016}
{Valenti}, S., {Howell}, D.~A., {Stritzinger}, M.~D., {et~al.} 2016,
  \href{http://dx.doi.org/10.1093/mnras/stw870}{\JournalTitle{\mnras}, 459,
  3939}

\bibitem[{van~der Walt {et~al.}(2011)van~der Walt, Colbert, \&
  Varoquaux}]{der_walt_2011_aa}
van~der Walt, S., Colbert, S.~C., \& Varoquaux, G. 2011,
  \href{http://dx.doi.org/10.1109/MCSE.2011.37}{\JournalTitle{Computing in
  Science Engineering}, 13, 22}

\bibitem[{Wolf {et~al.}(2017)Wolf, Bauer, \& Schwab}]{MesaScript}
Wolf, B., Bauer, E.~B., \& Schwab, J. 2017, wmwolf/MesaScript: A DSL for
  Writing MESA Inlists

\bibitem[{Wolf \& Schwab(2017)}]{MesaReader}
Wolf, B., \& Schwab, J. 2017, wmwolf/py\_mesa\_reader: Interact with MESA
  Output

\bibitem[{{Wongwathanarat} {et~al.}(2015){Wongwathanarat}, {M{\"u}ller}, \&
  {Janka}}]{Wongwathanarat2015}
{Wongwathanarat}, A., {M{\"u}ller}, E., \& {Janka}, H.-T. 2015,
  \href{http://dx.doi.org/10.1051/0004-6361/201425025}{\JournalTitle{\aap},
  577, A48}

\bibitem[{{Woosley} \& {Weaver}(1988)}]{Woosley1988}
{Woosley}, S.~E., \& {Weaver}, T.~A. 1988,
  \href{http://dx.doi.org/10.1016/0370-1573(88)90037-3}{\JournalTitle{\physrep},
  163, 79}

\end{thebibliography}

\end{document}